\documentclass{emulateapj}
\usepackage{amsmath}
\usepackage{color}
\usepackage{ulem}

\shorttitle{The VMC Survey XXVII. Young Stellar Structures in the LMC's Bar Star-Forming Complex}
\shortauthors{Sun et al.}
\slugcomment{Submitted for publication in The Astrophysical Journal}

\begin{document}

\title{The VMC Survey XXVII. Young Stellar Structures in the LMC's Bar Star-Forming Complex}

\author{Ning-Chen~Sun$^{1, 2}$, 
Richard~de~Grijs$^{1, 2, 3}$, 
Smitha~Subramanian$^{1}$, 
Kenji~Bekki$^{4}$, 
Cameron~P.~M.~Bell$^{5}$, 
Maria-Rosa~L.~Cioni$^{5, 6, 7}$, 
Valentin~D.~Ivanov$^{8, 9}$, 
Marcella~Marconi$^{10}$, 
Joana~M.~Oliveira$^{11}$, 
Andr\'es~E.~Piatti$^{12, 13}$, 
Vincenzo~Ripepi$^{10}$,
Stefano~Rubele$^{14, 15}$, 
Ben~L.~Tatton$^{11}$, 
Jacco~Th.~van~Loon$^{11}$}
\affil{
$^1$Kavli Institute for Astronomy and Astrophysics, Peking University, Yi He Yuan Lu 5, Hai Dian District, Beijing 100871, China; sunnc@foxmail.com, grijs@pku.edu.cn \\
$^2$Department of Astronomy, Peking University, Yi He Yuan Lu 5, Hai Dian District, Beijing 100871, China \\
$^3$International Space Science Institute -- Beijing, 1 Nanertiao, Hai Dian District, Beijing 100190, China \\
$^4$ICRAR, M468, The University of Western Australia 35 Stirling Highway, Crawley Western Australia, 6009, Australia \\
$^5$Leibniz-Institut f\"ur Astrophysik Potsdam, An der Sternwarte 16, Potsdam 14482, Germany \\
$^6$Universit\"at Potsdam, Institut f\"ur Physik und Astronomie, Karl-Liebknecht-Str. 24/25, Potsdam 14476, Germany \\
$^7$University of Hertfordshire, Physics, Astronomy and Mathematics, College Lane, Hatfield AL10 9AB, UK \\
$^{8}$ESO European Southern Observatory, Ave. Alonso de Cordova 3107, Casilla 19, Chile \\
$^{9}$ESO Garching: ESO,  Karl-Schwarzschild-Str. 2, 85748 Garching bei M\"unchen, Germany  \\ 
$^{10}$INAF-Osservatorio Astronomico di Capodimonte, Salita Moiariello, 16, I-80131 Napoli, Italy \\
$^{11}$School of Chemical \& Physical Sciences, Lennard-Jones Laboratories, Keele University, ST5 5BG, UK \\
$^{12}$Observatorio Astron\'omico, Universidad Nacional de C\'ordoba, Laprida 854, 5000, C\'ordoba, Argentina \\
$^{13}$Consejo Nacional de Investigaciones Cient\'ificas y T\'ecnicas, Av. Rivadavia 1917, C1033AAJ Buenos Aires, Argentina \\
$^{14}$Osservatorio Astronomico di Padova -- INAF, vicolo dell'Osservatorio 5, Padova I-35122, Italy \\
$^{15}$Dipartimento di Fisica e Astronomia, Universit\`a di Padova, vicolo dell'Osservatorio 2, Padova I-35122, Italy}

\begin{abstract}

Star formation is a hierarchical process, forming young stellar structures of star clusters, associations, and complexes over a wide scale range. The star-forming complex in the bar region of the Large Magellanic Cloud is investigated with upper main-sequence stars observed by the VISTA Survey of the Magellanic Clouds. The upper main-sequence stars exhibit highly non-uniform distributions. Young stellar structures inside the complex are identified from the stellar density map as density enhancements of different significance levels. We find that these structures are hierarchically organized such that larger, lower-density structures contain one or several smaller, higher-density ones. They follow power-law size and mass distributions as well as a lognormal surface density distribution. All these results support a scenario of hierarchical star formation regulated by turbulence. The temporal evolution of young stellar structures is explored by using subsamples of upper main-sequence stars with different magnitude and age ranges. While the youngest subsample, with a median age of log($\tau$/yr)~=~7.2, contains most substructure, progressively older ones are less and less substructured. The oldest subsample, with a median age of log($\tau$/yr)~=~8.0, is almost indistinguishable from a uniform distribution on spatial scales of 30--300~pc, suggesting that the young stellar structures are completely dispersed on a timescale of $\sim$100~Myr. These results are consistent with the characteristics of the 30~Doradus complex and the entire Large Magellanic Cloud, suggesting no significant environmental effects. We further point out that the fractal dimension may be method-dependent for stellar samples with significant age spreads.

\end{abstract}


\section{Introduction}

It has been suggested that star formation is a hierarchical process, forming young stellar structures over a wide range of scales \citep{Efremov1998, Elmegreen2000, Elmegreen2011, Gusev2014, Gouliermis2015, Gouliermis2017, Sun2017}. These structures include, for increasing size and decreasing density, star clusters, associations, and complexes. Studies of local star-forming regions \citep[e.g.][]{Gomez1993, Larson1995, Simon1997, Kraus2008} and nearby galaxies \citep[e.g.][]{Elmegreen2001, Elmegreen2006, Gouliermis2010, Gouliermis2015, Gouliermis2017} suggest that the young stellar structures display a high degree of substructuring and fractal properties, which may be inherited from the natal gas from which they form. After birth, the young stellar structures evolve rapidly toward uniform distributions before they are completely dispersed \citep{Gieles2008, Bastian2009, Gouliermis2015}. Beyond this simple picture, however, more exploration is needed to fully understand their properties, formation, evolution, and especially the roles played by related physical processes (e.g. gravity, turbulence, galactic dynamics, etc.).

Stellar complexes are important nurseries of new stars. They usually have kiloparsec scales and contain smaller young stellar structures such as associations and aggregates, which themselves are subclustered into compact star clusters \citep{Efremov1995}. Star formation in stellar complexes is not only influenced by global galactic properties, e.g. bars and spiral arms \citep{Binney1998}, but also regulated by local processes, e.g. gravity, turbulence, magnetic field, and stellar feedback \citep{MacLow2004}. On the other hand, the Large and Small Magellanic Clouds (LMC and SMC) are close neighbors of the Milky Way at distances of 50 and 60~kpc, respectively \citep{deGrijs2014, deGrijs2015}. The LMC is the prototype of barred Magellanic spiral galaxies. It has a flat stellar disk, a single-looping spiral arm, a stellar bar which is off-centered from the galaxy's dynamical center, and a large star-forming complex at the northwestern end of the bar \citep{Wilcots2009}. The SMC, however, is a late-type dwarf galaxy, with an elongated, cigar-shaped structure seen edge-on \citep{DOnghia2016}. They show signatures of interactions with one another as well as with the Milky Way's gravitational potential and halo gas \citep{deBoer1998, DOnghia2016, Belokurov2017, Subramanian2017}. Both Clouds exhibit active past and ongoing star formation \citep{Harris2004, Harris2009, Oliveira2009, Rubele2012, Rubele2015}. As a result, stellar complexes in the Magellanic Clouds provide unique laboratories for understanding young stellar structures.

\citet{Bastian2009} studied stellar structures in the LMC. Based on young massive (OB) stars, they found that their identified stellar groups have no characteristic length scale and exhibit a power-law luminosity function with index $-$2. Using stellar subsamples of different ages, they showed that while stars are born with a high degree of substructuring, older subsamples are progressively less clumpy, reaching a uniform distribution at $\sim$175~Myr. \citet{Bonatto2010} investigated the spatial correlation of star clusters and ``non-clusters" (which are basically nebula complexes and stellar associations) in the Magellanic Clouds. Using two-point correlation functions (TPCFs), they found that young star clusters present a high degree of spatial correlation with themselves and with non-clusters, which does not occur for old star clusters. They also noticed that star clusters (young and old) and non-clusters all have power-law size distributions but with different slopes. In general, these results are in agreement with the scenario referred to above.

Both studies focused on the global population of young stellar structures across the entire LMC (and SMC). Thus, they do not reveal whether there is any environmental dependence in the properties, formation, and evolution of the young stellar structures. It is possible to explore this issue by studying individual stellar complexes in the LMC. In \citet{Sun2017}, we reported the hierarchical patterns of the young stellar structures in the 30~Doradus-N158-N159-N160 complex (30 Dor complex hereafter) in the LMC. The structures were identified based on  upper main-sequence (upper-MS) stars observed by the VISTA Survey of the Magellanic Clouds \citep[VMC;][]{Cioni2011}. The results suggest that the 30~Dor complex is highly substructured in a scale-free manner, supporting the scenario of hierarchical star formation from a turbulent interstellar medium (ISM). The derived projected fractal dimension, $D_2$~=~1.6~$\pm$~0.3, is consistent with those of Galactic star-forming regions and NGC~346 in the SMC. Thus, no significant environmental dependence was discovered in the fractal dimension of the 30~Dor complex.

\begin{figure}
\centering
\includegraphics[scale=0.55,angle=0]{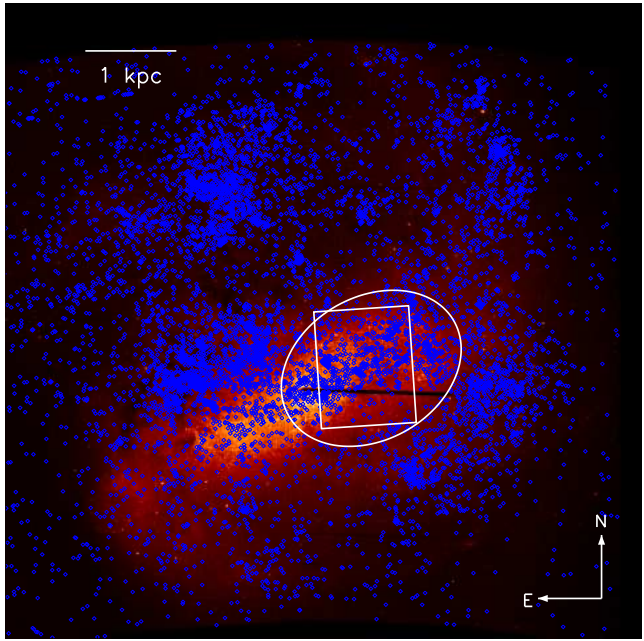}
\caption{{\it Colorscale}: density map of all stars with $V$~$<$~20~mag from the Magellanic Cloud Photometric Survey \citep[MCPS;][]{Zaritsky2004}. We obtain the map by simple star counts in bins of 1$\arcmin\times$1\arcmin. The nearly horizontal black line is caused by slight gaps between scans in their observations. The LMC bar corresponds to the northeast-southwest elongated structure with prominently high densities in this map. {\it Blue points}: MCPS stars with $V$~$<$~14.5~mag and $B-V$~$<$~0.5~mag, which are young and massive stars and trace recent star formation \citep{Bastian2009}. At the northwestern end of the bar, there is a concentration of such bright and blue stars, corresponding to the bar complex. Its approximate extent is indicated by the white ellipse. Note the other end of the bar contains significantly fewer bright and blue stars. The white rectangle shows the extent of VMC tile LMC~6\_4, which covers the major part of the bar complex. The map is centered R.A.(J2000)~=~05$^{\rm h}$18$^{\rm m}$48$^{\rm s}$, Dec.(J2000)~=~$-$68$^{\rm o}$42$\arcmin$00$\arcsec$. }
\label{lmc.fig}
\end{figure}

In this paper, we carry out a similar study of the stellar complex at the northwestern end of the LMC bar, which we shall refer to as the bar complex. A major part of this complex is covered by VMC tile LMC~6\_4, which is used for our analysis in this work (Fig.~\ref{lmc.fig}; see Section~\ref{vmc.sec} for the definition of a tile). The bar complex is an important component of the LMC. As an active star-forming nursery, it is abundant in molecular gas, H~{\scriptsize II} regions, young stellar objects, and star clusters \citep[e.g.][]{Fukui2008, Carlson2012, Piatti2015}. The previous studies of \citet{Bastian2009} and \citet{Bonatto2010}, however, have used small samples or subsamples, which contain several hundreds or thousands of objects. As a result, they do not provide sufficient spatial sampling to resolve the detailed inner structures of the bar complex. On the other hand, perturbations from the bar (\citealt{Gardiner1998}; but see \citealt{Harris2009} for a discussion against a dynamical bar) may have an influence on the bar complex; compared with the galaxy outskirts, it is less affected by external tidal forces \citep[e.g.][]{Fujimoto1990, Bekki2007, DOnghia2016}. It still remains unexplored whether these environmental processes cause any difference in the properties, formation, and evolution of the young stellar structures in the bar complex.

The goal of this paper is to understand the properties, formation, and evolution of young stellar structures in the bar complex. Using wide color and magnitude cuts, we construct a sample of more than 2.5~$\times$~10$^4$ upper-MS stars younger than $\sim$1~Gyr, an order of magnitude more than the previous samples mentioned above. The sample provides sufficient spatial sampling, allowing us to identify young stellar structures on scales $\gtrsim$10~pc. We then analyze the properties of these structures and discuss the effects of physical processes associated with their formation. We also use TPCFs to study the degree of substructuring in upper-MS subsamples of different magnitude and age ranges, which will help demonstrate any evolution of the young stellar structures. We discuss the environmental effects through comparisons with other regions or with the entire LMC.

This paper is organized as follows. Section~\ref{vmc.sec} describes the data used in this work. Our selection of upper-MS stars and their spatial distributions are outlined in Section~\ref{sample.sec}. Young stellar structures are identified and analyzed based on the full sample in Section~\ref{yss.sec}, while in Section~\ref{evo.sec} we use subsamples of upper-MS stars to explore their temporal evolution. Finally, we complete this paper with a summary and conclusions.

\section{Data}
\label{vmc.sec}

Data used in this work are from the VMC survey \citep{Cioni2011}, which is carried out with the Visible and Infrared Survey Telescope for Astronomy \citep[VISTA;][]{Sutherland2015}. The VMC survey is a multi-epoch, uniform, and homogeneous photometric survey of the Magellanic System (LMC, SMC, Bridge, and Stream). It uses the near-infrared $Y$, $J$, and $K_s$ bands, and the typical spatial resolution is $\sim$1$\arcsec$ or better. Because there are gaps between its 16 detectors, a sequence of six offsets is needed to observe a contiguous area of sky; the combined image is then referred to as a {\it tile}. Each tile covers an area of $\sim$1.5~deg$^2$ and is designed to be observed at three epochs in the $Y$ and $J$ bands and at 12 epochs in the $K_s$ band, corresponding to total exposure times of 2400~s, 2400~s, and 7500~s, respectively. The saturation limits are usually $Y$~=~12.9~mag, $J$~=~12.7~mag, and $K_s$~=~11.4~mag; typical 5$\sigma$ magnitude limits are $Y$=~21.9~mag, $J$~=~22.0~mag, and $K_s$~=~21.5~mag for stacked observations combining all epochs. Note, however, that these limits may vary with source crowding and sky conditions. The VMC survey is still ongoing, and is expected to cover 170~deg$^2$ on completion within its $\sim$9 years of observations. The star-forming complex analyzed in this work is based on tile LMC~6\_4. We retrieved the data of this tile as part of VMC Data Release 3 from the VISTA Science Archive (VSA). The VSA and the VISTA data flow pipeline are described by \citet{Cross2012} and \citet{Irwin2004}, respectively. We use point-spread function (PSF) photometry obtained with PSF-homogenized, stacked images from different epochs \citep[their Appendix~A]{Rubele2015}; the photometric errors and local completeness have also been calculated with artificial star tests \citep{Rubele2012, Rubele2015}.

The ``top" half of detector \#16 of the VISTA infrared camera (VIRCAM) has worse signal-to-noise ratios, since its pixel-to-pixel quantum efficiency varies on short timescales, leading to inaccurate flat-fields. This affects the southwestern corner of the analyzed tile, but it will be shown that the bar complex does not overlap with this region (Section~\ref{space.sec}). Thus we do not attempt to deal with this effect.

\section{The Upper-MS Sample}
\label{sample.sec}

\subsection{Sample Selection}
\label{ums.sec}

\begin{figure*}
\centering
\begin{tabular}{cc}
\includegraphics[scale=0.70,angle=0]{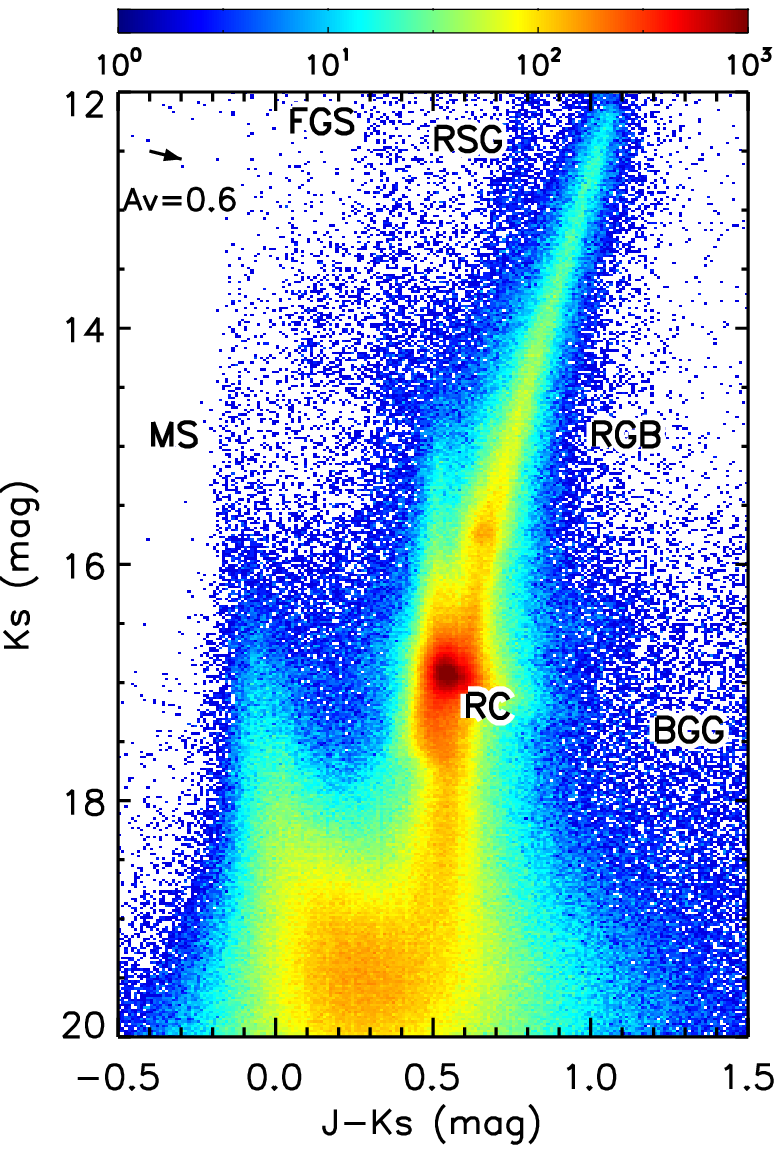}
\includegraphics[scale=0.70,angle=0]{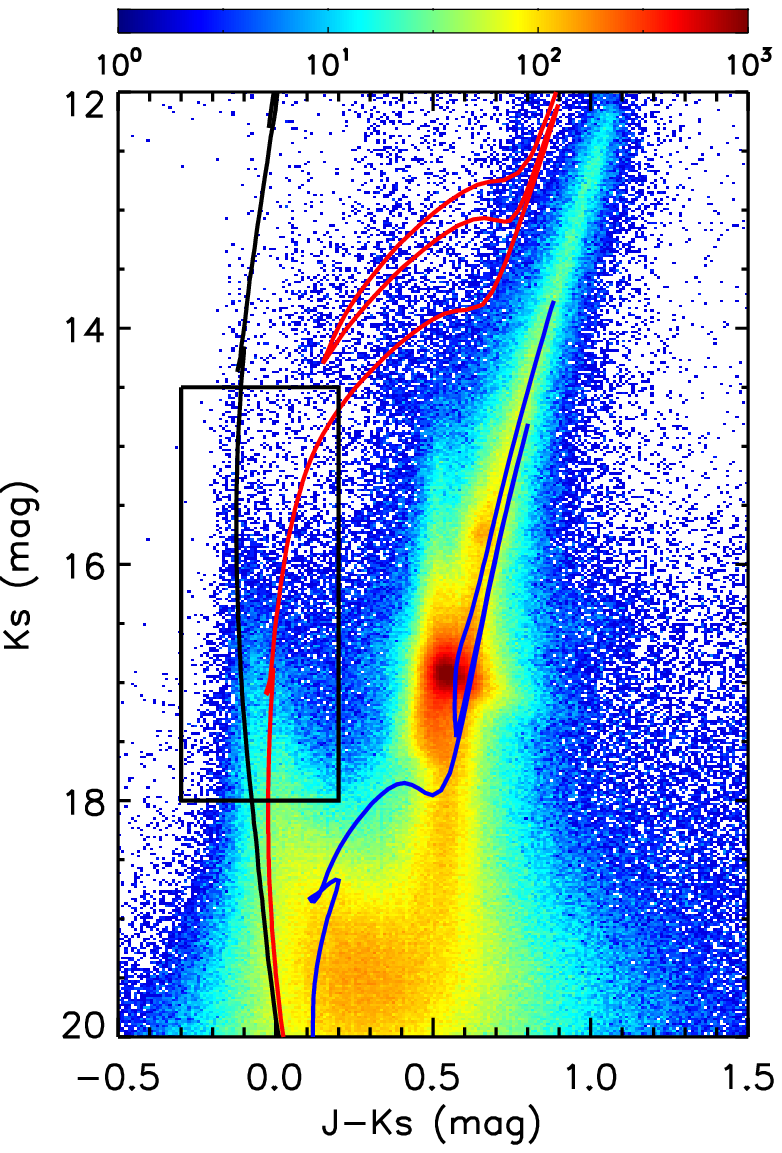}
\end{tabular}
\caption{($J - K_s$, $K_s$) color--magnitude Hess diagram of sources in tile LMC~6\_4. The color scales show the numbers of stars in each color--magnitude bin, with a bin size of 0.01~mag in color and 0.02~mag in magnitude. Main features are labeled in the left-hand panel, with ``FGS" and ``BGG" short for foreground Galactic stars and background galaxies, respectively. PARSEC (version 1.2S) isochrones of metallicity [Fe/H]~=~$-$0.3~dex and ages log($\tau$/yr)~=~7.0 (black line), 8.0 (red line), and 9.0 (blue line) are overplotted in the right-hand panel, shifted by a distance modulus of ($m$~$-$~$M$)$_0$~=~18.49~mag and an extinction of $A_V$~=~0.6~mag. The reddening vector is indicated by the arrow in the left-hand panel, and the black box in the right-hand panel shows the criteria adopted for selecting upper-MS stars. The color scales in both panels are identical; isochrones and selection criteria are not plotted in the left-hand panel simply for clarity.}
\label{cmd.fig}
\end{figure*}

Figure~\ref{cmd.fig} shows the ($J - K_s$, $K_s$) color--magnitude diagram (CMD) of stars in tile LMC~6\_4. The MS is clearly visible at colors $-$0.3~$<$~($J - K_s$)~$<$~0.2~mag. The red, populous branch is the red--giant branch (RGB), overlapping with the red clump (RC) at ($J - K_s$)~=~0.5~mag and $K_s$~=~17.0~mag. Slightly bluer than the RGB, red supergiants (RSG) can be seen at $K_s$~$<$~13.0~mag. The vertical strip at ($J - K_s$)~=~0.35~mag arises from foreground Galactic stars, and sources redder than ($J - K_s$)~=~1.0~mag are essentially background galaxies. In addition, an overdensity of stars can also be found around ($J - K_s$)~=~0.1~mag and 13.5~$<$~$K_s$~$<$~14.5~mag, most of which are primarily blue-loop stars.

In the right-hand panel of Fig.~\ref{cmd.fig}, PARSEC isochrones \citep[version 1.2S;][]{Bressan2012} of metallicity [Fe/H]~=~$-$0.3~dex \citep[which is typical for massive LMC stars;][]{Rolleston2002} and ages log($\tau$/yr)~=~7.0, 8.0, and 9.0 are overplotted. Offsets of 0.026~mag in $J$ and 0.003~mag in $K_s$ are subtracted from the isochrones to convert model magnitudes from the Vega system to the VISTA system \citep[for details, see][]{Rubele2012, Rubele2015}. The isochrones are shifted by the LMC's distance modulus of ($m$~$-$~$M$)$_0$~=~18.49~mag \citep{Pietrzynski2013, deGrijs2014} and then reddened by an extinction of $A_V$~=~0.6~mag. This extinction is found by matching the isochrone of log($\tau$/yr)~=~7.0 (whose MS extends the full magnitude range in the CMD) to the bluest edge of the MS at $K_{\rm s}$~$<$~18~mag. We apply the extinction coefficients $A_J$/$A_V$~=~0.283 and $A_{K_s}$/$A_V$~=~0.114, computed from the \citet{Cardelli1989} extinction curve with $R_V$~=~3.1 \citep{Girardi2008}.

Stars in the upper-MS are relatively young given that higher-mass stars have shorter MS lifetimes than their lower-mass counterparts. Indeed, stars in the upper-MS brighter than $K_s$~=~18.0~mag should be primarily younger than 1~Gyr, as indicated by the theoretical isochrones; thus, we select the upper-MS stars brighter than $K_s$~=~18.0~mag for the analysis in the following sections. We use a color window, $-$0.3~$<$~($J - K_{\rm s}$)~$<$~0.2~mag, to distinguish the upper-MS stars from the RGB and RC stars. The interstellar extinction in this region shows significant spatial variations, leading to the broad width of the upper-MS. Very high extinctions may shift stars out of the color window. However, such cases should be rare, since the adopted color window has a wide range in ($J - K_{\rm s}$) \footnote{For instance, an additional extinction of $A_V$~=~1.8~mag is needed to move a star from the log($\tau$/yr)~=~7.0 isochrone out of the color window. However, such an extinction is rare in the bar region \citep[see e.g.][]{Haschke2011}.}. An upper magnitude limit of $K_{\rm s}$~=~14.5~mag is applied to avoid the blue-loop stars. While there may be some contamination of blue-loop stars fainter than $K_{\rm s}$~=~14.5~mag, their number is small compared with the upper-MS stars; because of the rapid evolutionary phase, they will not spend much time there. The selection of upper-MS stars is indicated by the box in Fig.~\ref{cmd.fig} (right-hand panel), and the final upper-MS sample contains 25,232 stars in total.

It has long been known that young stellar structures are transient structures, except for bound star clusters on small scales, which may survive for a significant period \citep{Elmegreen1998}. They are usually younger than several hundred million years \citep{Efremov1995, Gieles2008, Bastian2009}. Thus, the upper-MS sample includes many stars which are older than the typical age of young stellar structures (see also Section~\ref{age.sec}). Despite this potential contamination, the young stellar structures can still be revealed with the upper-MS sample. It will be shown that old stars ($\gtrsim$100~Myr) in the sample are almost uniformly distributed (Section~\ref{evo.sec}); as a result, the surface density enhancements in the stellar distribution, which are identified as young stellar structures (Section~\ref{kde.sec}), are dominated only by young stars. On the other hand, although it is possible to reduce the contamination by adopting a brighter upper magnitude limit, this will lead to a very small sample with poor spatial sampling insufficient to resolve the small-scale young stellar structures. Moreover, the upper-MS sample containing both young and old stars allows us to investigate their temporal evolution (Section~\ref{evo.sec}). Thus, we use this upper-MS sample for the following analysis.

\subsection{Spatial Distributions}
\label{space.sec}

\begin{figure*}
\centering
\begin{tabular}{cc}
\includegraphics[scale=0.75,angle=0]{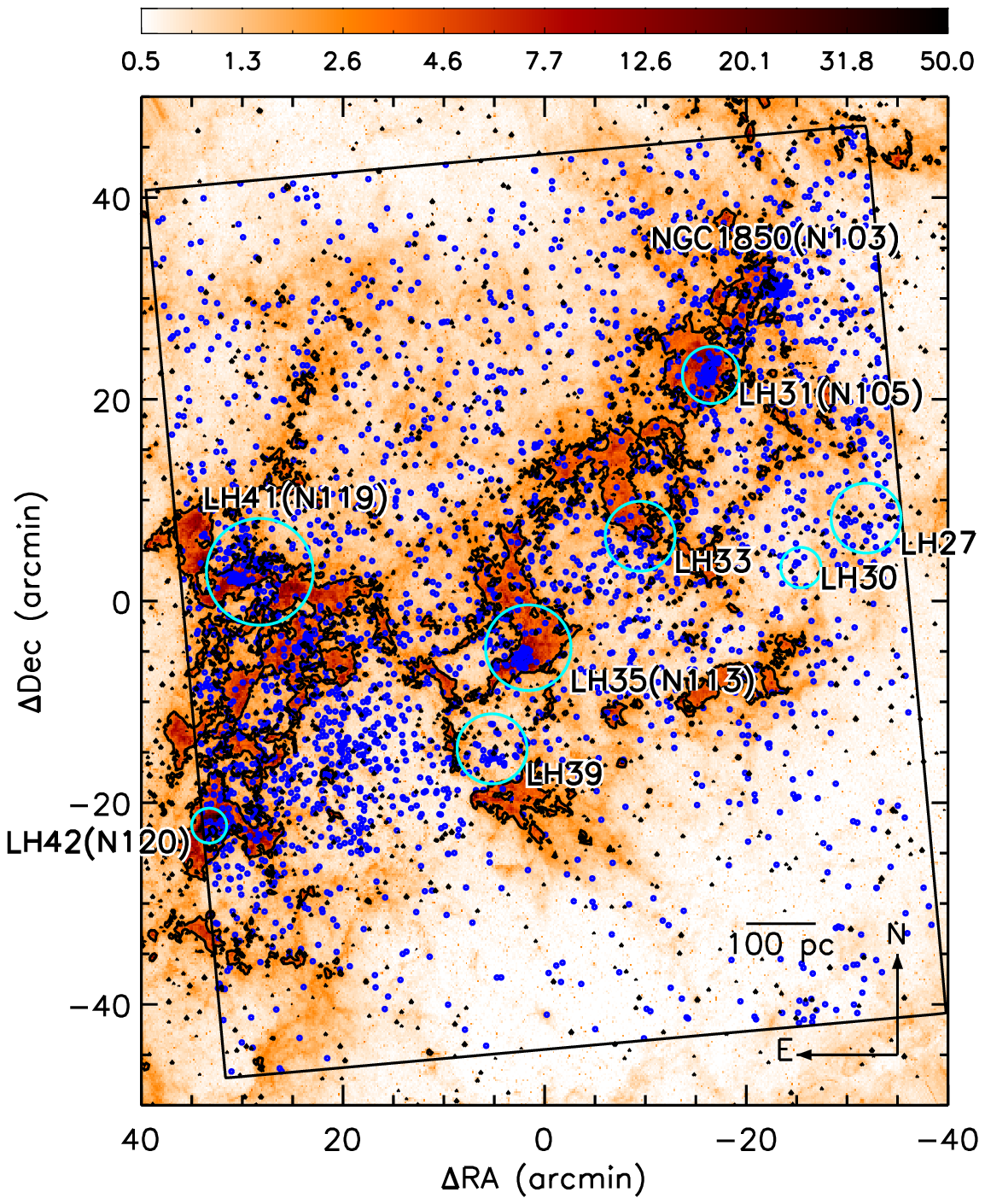}
\includegraphics[scale=0.75,angle=0]{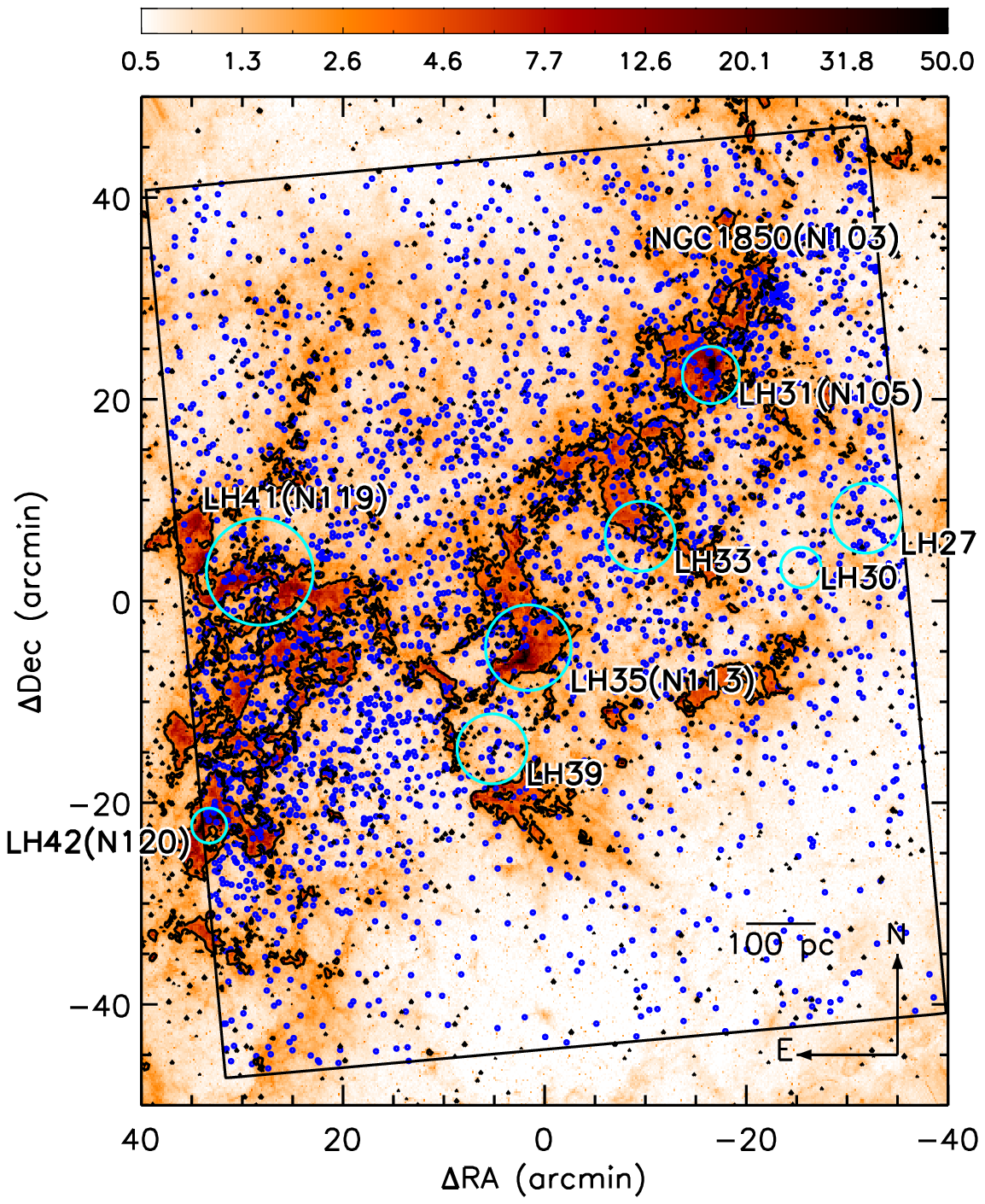}
\end{tabular}
\caption{Spatial distributions of the upper-MS stars (blue points) in the magnitude ranges of 14.5~$<$~$K_s$~$<$~16.5~mag (left-hand panel) and 16.5~$<$~$K_s$~$<$~18.0~mag (right-hand panel). There are 2827 stars with 14.5~$<$~$K_s$~$<$~16.5~mag, which are all displayed in the left-hand panel. The same number of stars are randomly selected and displayed in the right-hand panel from the 22,403 stars with 16.5~$<$~$K_s$~$<$~18.0~mag. This is simply for clarity and for the ease of comparison between the two panels. The background colorscale is the Spitzer/IRAC~8.0~$\mu$m dust emission map from SAGE. The color bar is in units of MJy~sr$^{-1}$; contours of 3 MJy~sr$^{-1}$ are also overplotted. Eight LH associations are labeled, with the cyan circles showing their approximate extents; the circles' radii are calculated based on the associations' geometric means of the major and minor axes given by \citet{Lucke1970}. If an LH association is associated with a \citet{Henize1956} nebula, the nebula name is also labeled in the bracket after the LH designation. To the northwest of LH31 is the young, populous star cluster NGC~1850. The (0, 0) position corresponds to R.A.(J2000)~=~05$^{\rm h}$12$^{\rm m}$55$^{\rm s}$.5, Dec.(J2000)~=~$-$69$^{\rm o}$16$\arcmin$41$\arcsec$, and the black rectangle shows the extent of VMC tile LMC~6\_4.}
\label{ums.fig}
\end{figure*}

Figure~\ref{ums.fig} shows the spatial distribution of the selected upper-MS stars. Simply for clarity, we show stars brighter and fainter than $K_s$~=~16.5~mag in two panels separately. All 2827 stars brighter than $K_s$~=~16.5~mag are displayed in the left-hand panel, while the same number of stars are randomly selected and displayed in the right-hand panel from the 22,405 stars fainter than $K_s$~=~16.5~mag. We do not show all the stars fainter than $K_s$~=~16.5~mag to avoid symbols crowding in the figure. From the left-hand panel, it is apparent that the distribution of the upper-MS stars brighter than $K_s$~=~16.5~mag is not uniform but highly clumpy and substructured. Compared with a dispersed stellar distribution across the tile, many of the stars reside in groups with high surface densities. These groups correspond well with eight \citet[LH;][]{Lucke1970} associations, the positions and extents of which are labeled in the figure. Some other groups not cataloged by \citet{Lucke1970} are also visible; for instance, the stellar group to the northwest of LH~31 corresponds primarily to the young, populous star cluster NGC~1850 \citep{Vallenari1994}. Another large concentration of upper-MS stars can also be found between LH~39, LH~41, and LH~42.

The right-hand panel shows the spatial distribution of stars fainter than $K_s$~=~16.5~mag, which is less clumpy than that of their brighter counterparts. Some of the LH associations, e.g., LH~33 and LH~39, become less prominent and can be barely seen in the stellar distribution. Considering that the fainter stars have an older average age than the brighter ones, their different distributions may suggest the presence of an evolutionary effect. This will be discussed in detail in Section~\ref{evo.sec}.

Figure~\ref{ums.fig} also shows the IRAC~8.0~$\mu$m emission map from the Spitzer legacy program ``Surveying the Agents of Galaxy Evolution" \citep[SAGE;][]{Meixner2006}. The 8.0~$\mu$m emission comes mainly from hot dust, which is heated by young stars and re-radiated at infrared wavelengths. It can be seen that some of the stellar groups are associated with dust emission, for instance, NGC~1850, LH~31, LH~33, LH~35, LH~41, and LH~42. In contrast, LH~27, LH~30, and LH~39 are not associated with dust emission; and moreover, the large stellar concentration between LH~39, LH~41, and LH~42 lies in a void of dust emission, with the surrounding dust emission exhibiting a half-circular boundary. It is possible that these stars have dispersed the ISM through their radiation, stellar winds, and/or supernovae.

\citet{Henize1956} has cataloged H$\alpha$-emitting nebulae in the Magellanic Clouds, and this region is very abundant in such nebulae, which are primarily H~{\scriptsize II} regions, wind-driven shells, supernova remnants, etc., or complexes of different types (e.g. \citealt{Laval1992, AmbrocioCruz1998}; see also \citealt{Carlson2012}). Some of these nebulae are associated with the above-mentioned stellar groups, e.g. N103, N105, N113, N119, and N120, which are indicated in the figure. An extended nebula, N117, is associated with the stellar concentration between LH~39, LH~41, and LH~42. In addition, a number of Henize nebulae with small sizes are also located in this region \citep[see Fig. 6 of][]{Henize1956}. The presence of H$\alpha$-emitting nebulae suggests active feedback from the young stars to the ISM.

\section{Young Stellar structures}
\label{yss.sec}

\subsection{Identification and Dendrograms}
\label{kde.sec}

\begin{figure*}
\centering
\begin{tabular}{cc}
\includegraphics[scale=0.80,angle=0]{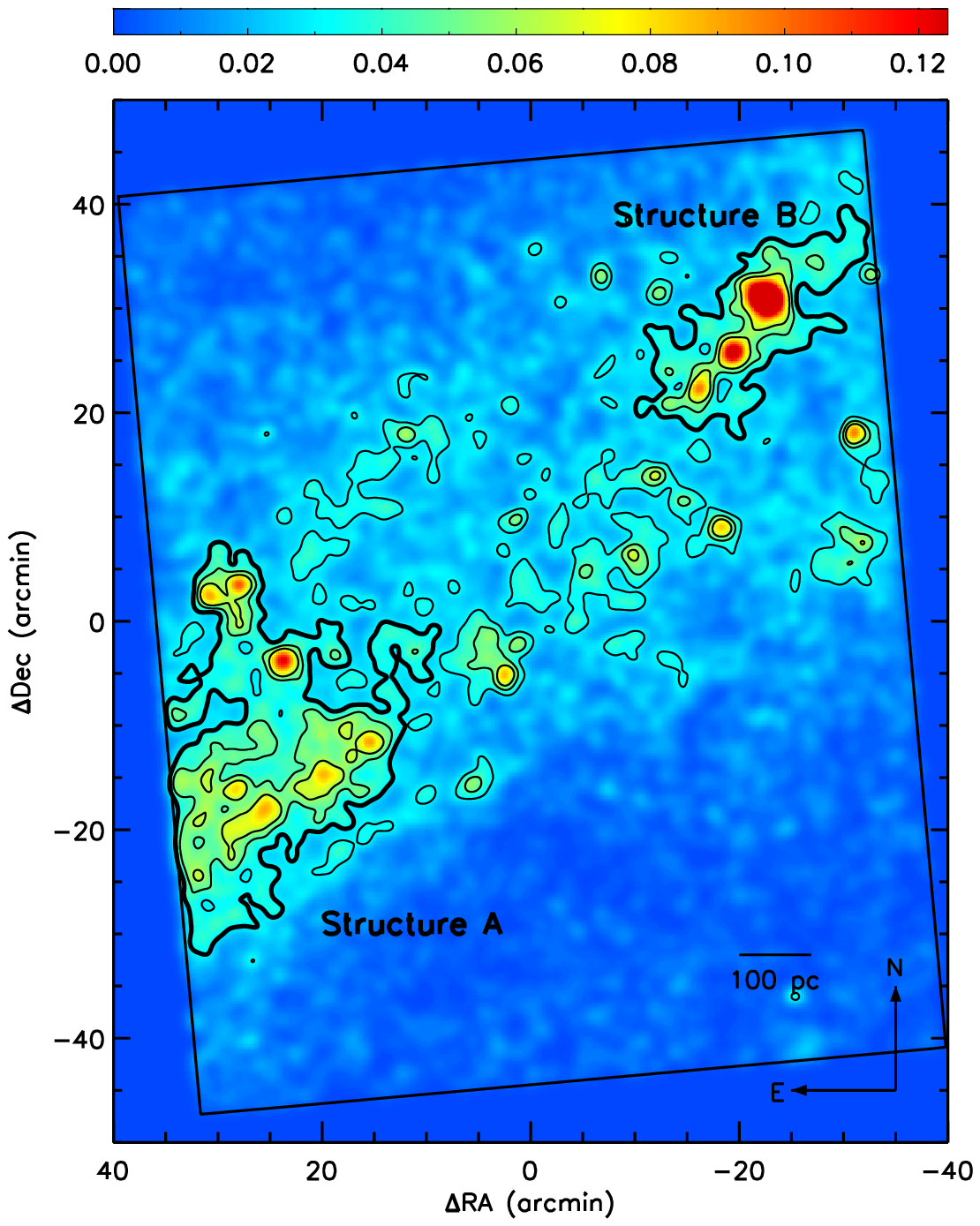}
\includegraphics[scale=0.70,angle=0]{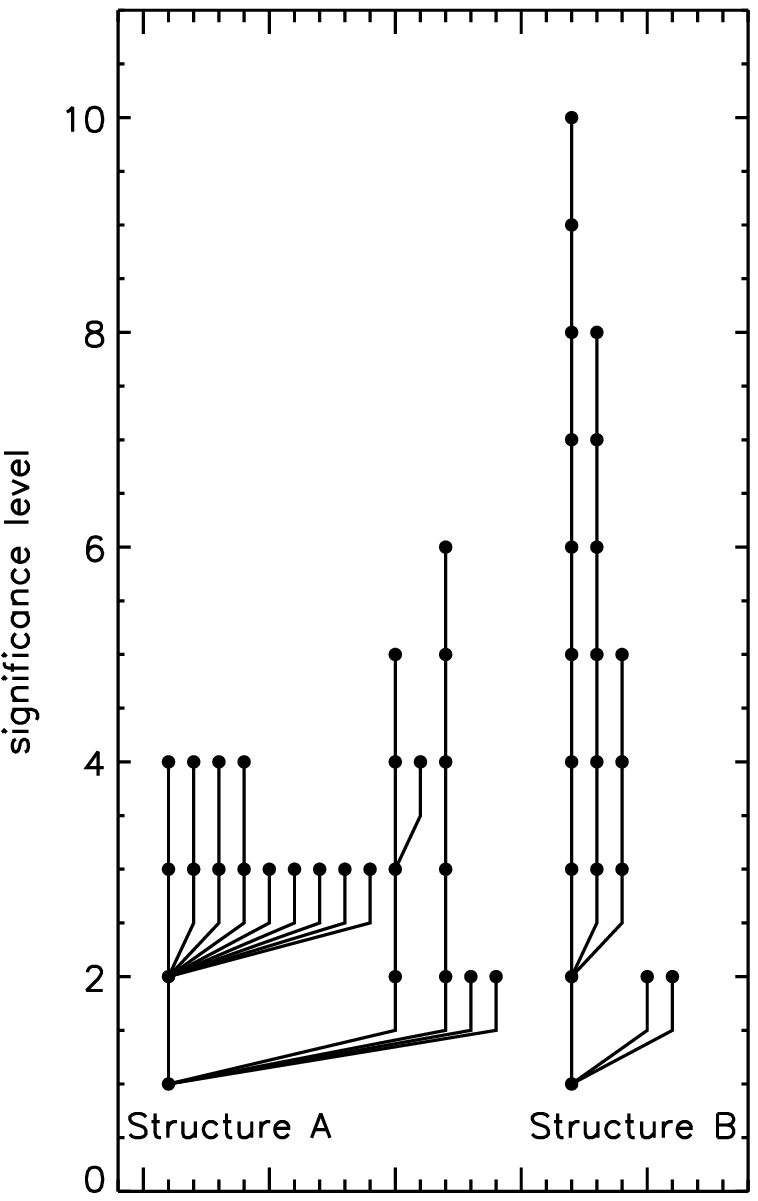}
\end{tabular}
\caption{Left: KDE map of the upper-MS stars. The color bar is in units of stars~pc$^{-2}$. The contours correspond to 1$\sigma$, 2$\sigma$, and 3$\sigma$ significance levels, and the thick-lined contours denote Structures~A and B. The (0, 0) position corresponds to R.A.(J2000)~=~05$^{\rm h}$12$^{\rm m}$55$^{\rm s}$.5, Dec.(J2000)~=~$-$69$^{\rm o}$16$\arcmin$41$\arcsec$, and the black rectangle shows the extent of VMC tile LMC~6\_4. Right: dendrograms of Structures~A and B.}
\label{kde.fig}
\end{figure*}

In the previous section, we have shown that the upper-MS stars exhibit large numbers of overdensities in the bar complex. In this paper, these overdensities are all referred to as young stellar structures, regardless of their mass or length scales, whether gravitationally bound or unbound. In order to quantitively analyze the young stellar structures in the bar complex, we need to identify them in a systematic way; to do this, we adopt the same method as detailed in \citet{Gouliermis2015, Gouliermis2017}. Firstly, we construct a surface density map of the upper-MS stars (Fig.~\ref{kde.fig}, left-hand panel) through kernel density estimation (KDE). This is done by convolving the map of the upper-MS stars with a Gaussian kernel. The choice of an optimal kernel width is best decided through experimentation \citep{Gouliermis2017}. The kernel width specifies the resolution of the KDE map, but small kernels lead to significant noise as well. Testing various kernel sizes shows that a standard-deviation width of 10~pc offers a good balance between resolution and noise. With this width it is possible to resolve structures of sizes comparable to or larger than 10~pc. We can achieve this resolution because the upper-MS sample provides sufficient spatial sampling based on its $\sim$2.5~$\times$~10$^4$ stars. Smaller samples will unnecessarily lead to poorer resolutions. The resultant KDE map has a median value of 0.011 stars~pc$^{-2}$, a mean value of 0.015 stars~pc$^{-2}$, and a standard deviation of 0.016 stars~pc$^{-2}$.

We then identify young stellar structures as stellar surface overdensities above the mean value, of different significance levels from 1$\sigma$ to 10$\sigma$, in steps of 1$\sigma$. To avoid spurious detections, we require that the iso-density contour of each structure should enclose at least $N_{\rm min}$~=~5 upper-MS stars. As a result, 52, 23, 20, 13, 6, 3, 2, 2, 1, and 1 structures are identified at the 10 levels of increasing significance, i.e., 123 structures in total. The catalog of identified structures, along with their physical parameters (see Sections~\ref{size.sec}--\ref{density.sec}), is given in Table~\ref{catalog.tab}. The choice of $N_{\rm min}$ is arbitrary, since larger values of $N_{\rm min}$ may miss out small real structures while smaller values lead to more spurious identifications. Changing $N_{\rm min}$ does not affect the conclusions in this section but may be important, and will be discussed, in the context of parameter statistics of the young stellar structures (Secions~\ref{size.sec}--\ref{density.sec}).

The KDE map shows that many of the young stellar structures exhibit very irregular morphologies; and moreover, the young stellar structures at different significance levels are organized in a hierarchical way. This is especially obvious for Structures~A and B, the two large-sized young stellar structures at 1$\sigma$ significance level located in the southeast and northwest of the field, respectively. Structures~A and B contain five and three substructures at 2$\sigma$ level, respectively; the substructures are smaller in size than their parent structures, and going up to even higher significance levels they may vanish, survive but contract in size, or fragment into several even smaller sub-substructures. These are illustrated with dendrograms -- structure trees showing the ``parent--child" relations of young stellar structures found at various significance levels (Fig.~\ref{kde.fig}, right-hand panel). This hierarchical subclustering of young stars has also been reported for a number of star-forming regions and galaxies \citep{Gouliermis2010, Gouliermis2015, Kirk2011, Gusev2014, Sun2017}, and is an indicator of hierarchical star formation over a range of length scales \citep{Efremov1995, Elmegreen2000, Elmegreen2011}.

\begin{deluxetable}{ccrrccc}
\tablecolumns{7}
\tablecaption{Identified Young Stellar Structures.\label{catalog.tab}}
\tablehead{
\colhead{ID} & \colhead{level} & \colhead{$\alpha$(J2000)} & \colhead{$\delta$(J2000)} & \colhead{$R$} &  \colhead{$N_\ast$} & \colhead{$\Sigma$}\\
\colhead{ } & \colhead{($\sigma$)} & \colhead{(deg)} & \colhead{(deg)} & \colhead{(pc)} & \colhead{ } & \colhead{(pc$^{-2}$)} \\
\colhead{(1)} & \colhead{(2)} & \colhead{(3)} & \colhead{(4)} & \colhead{(5)} & \colhead{(6)} & \colhead{(7)}
}
\startdata
   1 &  1 & 146.1157 & $-$67.8603 & 212.9 & 4419.2 & 0.03 \\
   2 &  1 &  53.9892 & $-$40.0755 & 139.9 & 1833.3 & 0.03 \\
   3 &  1 &  59.7185 & $-$63.4666 &  72.1 &  346.6 & 0.02 \\
   4 &  1 & 101.9242 & $-$52.8468 &  80.0 &  233.8 & 0.01 \\
   5 &  1 &  92.1920 & $-$72.2725 &  60.4 &  421.4 & 0.04 \\
   6 &  1 &  29.4799 & $-$51.5429 &  50.2 &  239.9 & 0.03 \\
   7 &  1 &  60.3419 & $-$55.3113 &  62.6 &  170.8 & 0.01 \\
   8 &  1 &  35.6166 & $-$45.5387 &  39.8 &  161.7 & 0.03 \\
   9 &  1 &  46.3646 & $-$56.6696 &  34.9 &  151.7 & 0.04 \\
  10 &  1 &  77.6561 & $-$65.8653 &  33.4 &   91.9 & 0.03 \\
  11 &  1 & 115.9339 & $-$57.2507 &  32.3 &  102.4 & 0.03 \\
  12 &  1 & 119.0661 & $-$82.0242 &  25.3 &   81.6 & 0.04 \\
  13 &  1 &  51.9121 & $-$69.6249 &  22.2 &   58.0 & 0.04 \\
  14 &  1 & 171.0498 & $-$72.8845 &  22.0 &   47.5 & 0.03 \\
  15 &  1 &  72.9400 & $-$60.0954 &  18.9 &   41.3 & 0.04
\enddata
\tablecomments{{\it Column~1}: ID number for each young stellar structure; {\it Column~2}: the significance level; {\it Columns~3}: the right ascension of the center of each young stellar structure, defined as $\alpha$~=~($\alpha_{\rm min}$~+~$\alpha_{\rm max}$)/2, where $\alpha_{\rm min}$ and $\alpha_{\rm max}$ are the minimum and maximum right ascension of the iso-density contour of each structure, respectively; {\it Column~4}: same as Column~3 but for the declination; {\it Columns~5-7}: parameters of size, mass (expressed in $N_\ast$), and surface density (see Sections~\ref{size.sec}-\ref{density.sec}). Only the first 15 records are shown for example. The complete catalog is available online.}
\end{deluxetable}

\subsection{Size Distribution}
\label{size.sec}

\begin{figure}
\centering
\includegraphics[scale=0.67,angle=0]{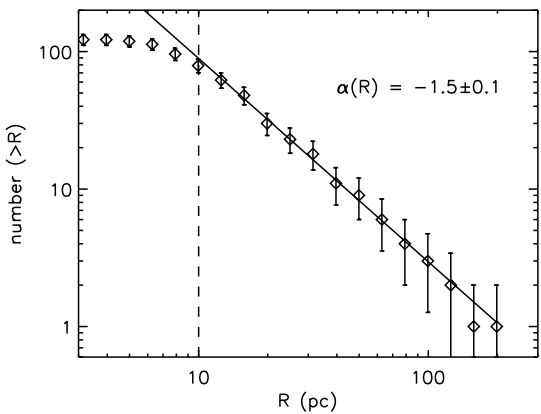}
\caption{Cumulative size distribution of young stellar structures. The vertical dashed line shows the threshold of completeness, and the solid lines is a power-law fit to the data above the threshold. The error bars reflect Poissonian uncertainties.}
\label{size.fig}
\end{figure}

The young stellar structures span a wide range of sizes, from the largest Structure~A, to the smallest ones, which are comparable to the kernel width. The size of each young stellar structure, $R$, is estimated based on the radius of a circle which has the same area as that covered by the iso-density contour of the structure. The results for all structures are listed in Table~\ref{catalog.tab}. Figure~\ref{size.fig} shows the cumulative size distribution. The distribution is approximately a single power law for larger sizes but shows significant flattening at smaller sizes. We note that there are two main factors affecting the completeness of young stellar structures (Section~\ref{kde.sec}). On the one hand, the KDE map, from which they have been identified, has resolution of 10~pc; thus, structures smaller than this size will probably be smeared out by the convolution. On the other hand, each structure has been required to contain at least $N_{\rm min}$~=~5 stars; as a result, small structures may not contain enough upper-MS stars at the less populated high-mass end of the stellar initial mass function (IMF). To assess the latter effect, we carried out an experiment similar to that of \citet{Sun2017}, i.e. by changing $N_{\rm min}$ to 10 and 15 and repeating the structure identification process. The resulting structure size distributions below 10~pc do indeed change with different values of $N_{\rm min}$; beyond 10~pc, however, the distributions remain unaffected (not shown). Considering all this, the young stellar structures are complete beyond $R$~=~10~pc. We fit a single power-law function to the size distribution above this value, which suggests a power-law slope of $\alpha(R)$~=~$-$1.5~$\pm$~0.1.

The sizes of the substructures inside a fractal follow
\begin{equation}
N(> R) \propto R^{-D};
\end{equation}
where $D$ is the fractal dimension \citep{Mandelbrot1983, Elmegreen1996}. Thus, the size distribution of the young stellar structures is consistent with a (projected) fractal dimension of $D_2$~=~1.5~$\pm$~0.1. The subscript `2' indicates that the young stellar structures have been identified and analyzed based on two-dimensional projections. It is not easy to obtain the three-dimensional (volume) fractal dimension, $D_3$. A relation, $D_3$~=~$D_2$~+~1, has been proposed by \citet{Beech1992}; however, this relation applies only when the perimeter-area dimension of the object's projection is the same as that of a slice \citep{Elmegreen2004}. The relation between $D_2$ and $D_3$ has also been investigated based on simulations \citep[e.g.][]{Sanchez2005, Gouliermis2014}.

There is a similarity between the ISM and the young stellar structures identified here. First, similar to the young stellar structures, the ISM also displays irregular morphologies and contains large amounts of substructures (clouds, clumps, cores, and filaments, etc.) which are hierarchically organized \citep{Rosolowsky2008}. Second, the ISM substructures also follow a power-law size distribution, which indicates a scale-free behavior \citep[e.g.][]{Elmegreen1996}. The third aspect of their similarity comes from the fractal dimension. The projected fractal dimension of the ISM has been investigated based on the perimeter-area relation of its projected boundaries. Typical values are close to $D_2$~=~1.4--1.5 (e.g. \citealt{Beech1987, Scalo1990, Falgarone1991, Vogelaar1994, Lee2004, Lee2016}; although smaller values have also been reported by e.g. \citealt{Dickman1990, Hetem1993}), which are consistent with the fractal dimension as derived for the young stellar structures. Using power-spectrum analysis, \citet{Stanimirovic1999, Stanimirovic2000} reported $D_2$~=~1.4 or 1.5 for the ISM in the SMC, also close to that of the young stellar structures. On the other hand, it is possible to measure the {\it volume} fractal dimension of the ISM, since clouds along the line of sight can be distinguished by their velocities. For instance, \citet{Elmegreen1996} reported $D_3$~=~2.3~$\pm$~0.3 based on the size distribution for a number of Galactic molecular clouds, and \citet{RomanDuval2010} found $D_3$~=~2.36~$\pm$~0.04 using the mass--size relation. If the relation $D_3$~=~$D_2$~+~1 holds for the ISM, these results would not be far from the fractal dimension of the young stellar structures. Unfortunately, there is no reported measurement of the fractal dimension of the ISM in the LMC bar region. However, it has been suggested that, despite a few exceptions, the fractal dimension is invariant from cloud to cloud, regardless of their nature as star-forming or quiescent, whether gravitationally bound or unbound \citep[e.g.][]{Williams2000}.

A fractal dimension of $D_3$~=~2.4 is consistent with laboratory results of numerical turbulent flows \citep{Sreenivasan1991, Elmegreen2004, Federrath2009}. As a result, turbulence has been argued to play a major role in creating the hierarchical structures in the ISM. In addition, agglomeration with fragmentation \citep{Carlberg1990} and self-gravity \citep{deVega1996} may also contribute \citep[see also][]{Elmegreen2000}. The similarity discussed in the previous paragraph suggests that the young stellar structures may have inherited the irregular morphologies, hierarchy, size distribution, and fractal dimension from the ISM substructures where they form. After birth, the young stellar structures are expected to evolve toward uniform distributions before they are finally dispersed \citep[Section~\ref{evo.sec}; see also][]{Gieles2008, Bastian2009, Gouliermis2015}. As a result, $D_2$ is expected to increase with age toward an ultimate value of 2. For the bar complex, the small value of $D_2$~=~1.5~$\pm$~0.1 suggests insignificant evolutionary effects in the structures' size distribution. The reason for this will be further discussed in Section~\ref{mix.sec}.

A power-law size distribution has also been found for the 30~Dor complex \citep{Sun2017} with a projected fractal dimension of 1.6~$\pm$~0.3, which is in agreement with the value found here for the bar complex, within errors. It has been proposed that star formation in the LMC is influenced by the perturbation of the off-center bar \citep{Gardiner1998} or by interactions with the SMC \citep[e.g.][]{Fujimoto1990, Bekki2007}; specifically, star formation in the 30~Dor complex may be induced by the bow shock as the LMC moves through the Milky Way's halo \citep{deBoer1998}. The 30~Dor and bar complexes are located in very different galactic environments; however, no significant difference in $D_2$ is found, considering the measurement uncertainties. Stars in other Galactic or SMC star-forming regions have $D_2$~=~1.4--1.5 \citep[e.g.][]{Larson1995, Simon1997, Gouliermis2014}, also close to that of the bar complex. On the other hand, galaxy-wide young stellar distributions exhibit a larger range of $D_2$ from $\sim$1.0 to $\gtrsim$1.8, which may reflect different clustering properties, evolutionary effects, or a method dependence \citep{Elmegreen2001, Elmegreen2006, Elmegreen2014, Gouliermis2015}. We refer the reader to \citet[][their Section~6]{Sun2017} for a more detailed discussion of comparisons of $D_2$.

\subsection{Mass Distribution and the Mass--Size Relation}
\label{mass.sec}

\begin{figure}
\centering
\includegraphics[scale=0.67,angle=0]{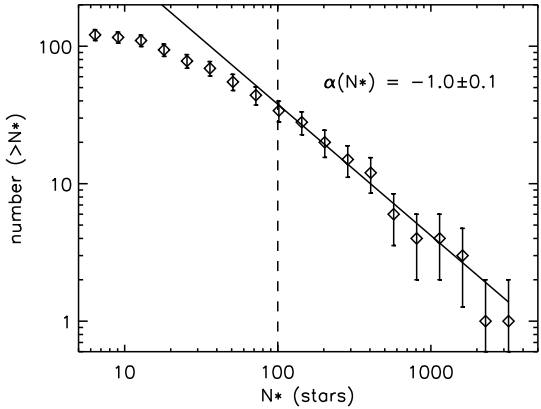}
\caption{Cumulative mass distribution of young stellar structures. The structure mass is represented by $N_\ast$ as defined by Equation~\ref{Nstar.eq}. The vertical dashed line shows the threshold of completeness, and the solid line is a power-law fit to the data above the threshold. The error bars reflect Poissonian uncertainties.}
\label{mass.fig}
\end{figure}

\begin{figure}
\centering
\includegraphics[scale=0.67,angle=0]{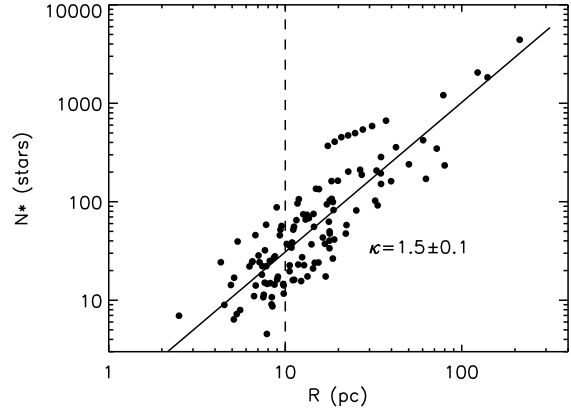}
\caption{Mass--size relation of all identified young stellar structures. The vertical dashed line at $R$~=~10~pc indicates the resolution of the KDE map.}
\label{masssize.fig}
\end{figure}

The number of upper-MS stars in a structure, $N_\ast$, provides an approximate representation of the structure mass, assuming that all structures have a similar age and that the stellar IMF is fully sampled. To obtain this quantity, we first correct the photometric incompleteness by assigning weights to the stars, defined as $w$~=~1.0/min[$f_J$, $f_{K_{\rm s}}$], where $f_J$ and $f_{K_{\rm s}}$ are the local completeness in the $J$ and $K_{\rm s}$ bands, respectively (Section~\ref{vmc.sec}). We then calculate $N_\ast$ with
\begin{equation}
\label{Nstar.eq}
N_\ast = \sum_i w_i - \Sigma_{\rm bg} A
\end{equation}
in which the subscript $i$ runs for all stars enclosed by the structure's iso-density contour, $\Sigma_{\rm bg}$~=~0.015~stars~pc$^{-2}$ is the mean surface density of the KDE map, and $A$ is the area of the iso-density contour (see Section~\ref{kde.sec} and Fig.~\ref{kde.fig}). We use the last term to estimate the number of background stars in chance alignment with the structure. Table~\ref{catalog.tab} lists $N_\ast$ for all the structures.

Figure~\ref{mass.fig} shows the cumulative distribution of $N_\ast$, which for brevity we shall refer to as the mass distribution. Beyond $N_\ast$~=~100~stars, the mass distribution is well described by a single power law; at $N_\ast$~$<$~100~stars, however, the mass distribution shows a deficiency of structures with respect to the extrapolation from the higher-mass end. Similarly to the size distribution, this is also caused by the incompleteness of young stellar structures. This is supported by Fig.~\ref{masssize.fig}, which shows a strong correlation of mass and size (with a Pearson correlation coefficient of 0.90). It is apparent that structures more massive than $N_\ast$~=~100~stars are all larger than 10~pc, while structures below this mass may be larger or smaller than 10~pc. Recall that the young stellar structures are complete beyond this size, the mass--size relation suggests that the completeness threshold lies at $N_\ast$~=~100~stars in the mass distribution.

Beyond $N_\ast$~=~100~stars, the cumulative mass distribution has a power-law slope of $\alpha(N_\ast)$~=~$-$1.0~$\pm$~0.1. This translates into a \textit{differential} mass function of the form $n(M)$d$M$~$\propto$~$M^{-\beta}$d$M$, with $\beta$~=~2.0~$\pm$~0.1. A mass function slope of 2 is predicted by the scenario of hierarchical star formation \citep{Fleck1996, Elmegreen2008}. This is in also agreement with studies of young stellar structures in NGC~628 \citep{Elmegreen2006}, M33 \citep{Bastian2007}, the LMC \citep{Bastian2009}, and the SMC \citep{Oey2004}. The same slope is also found for young star clusters in a wide variety of environments, e.g. the solar neighborhood \citep{Battinelli1994}, the Magellanic Clouds \citep{Hunter2003, deGrijs2006, deGrijs2008}, the Whirlpool Galaxy \citep{Bik2003}, and the extreme starbursts, NGC~3310 and NGC~6745 \citep{deGrijs2003}.

As mentioned in the first paragraph of this subsection, $N_\ast$ is proportional to mass if the young stellar structures have similar ages and if the stellar IMF is fully sampled. Age differences or stochastic sampling of the IMF may lead to uncertainties in the slope of the mass function. Note, however, that the slope is derived for structures with $N_\ast$~$\ge$~100~stars. As a result, the stochastic sampling effect is very small. On the other hand, we have assessed the $K_{\rm s}$-band luminosity functions (LFs) of the upper-MS stars which are located inside the iso-density contours of these young stellar structures (not shown). There are 34 structures with $N_\ast$~$>$~100~stars, and 28 of them have LFs with indistinguishable shapes. Thus, their age differences are not very significant. Only 7 out of the 34 structures have LFs with significantly different shapes, suggesting possible age differences. These structures have $N_\ast$ ranging from 103 to 358~stars; thus, they may only slightly affect the first few bins above $N_\ast$~=~100 in Fig.~\ref{mass.fig}. Considering all this, we suggest that possible age differences do not influence the derived mass function slope.

Returning to the mass--size relation (Fig.~\ref{masssize.fig}), the mass and size of young stellar structures follow a power-law correlation of $N_\ast$~$\propto$~$R^\kappa$, in which $\kappa$~=~1.5~$\pm$~0.1. Power-law mass--size relations with fractional slopes are also reported for young stellar structures in e.g. M33 \citep{Bastian2007}, NGC~6503 \citep{Gouliermis2015}, and NGC~1566 \citep{Gouliermis2017}. This power-law relation is expected from a fractal distribution of upper-MS stars \citep{Elmegreen1996}\footnote{Power-law mass--size relations do not necessarily come from fractal distributions. For example, stars following a centrally concentrated distribution of $\Sigma$~$\propto$~$R^{-0.5}$ (where $\Sigma$ is the local surface density and $R$ is the distance from the center) also produce the observed mass--size relation; but this distribution obviously does not apply to the upper-MS stars.}. Moreover, this power-law slope is another definition of the fractal dimension \citep{Mandelbrot1983}, which is consistent with that from the structure size distribution (Section~\ref{size.sec}).

\subsection{Density Distribution and the Density--Size Relation}
\label{density.sec}

\begin{figure}
\centering
\includegraphics[scale=0.67,angle=0]{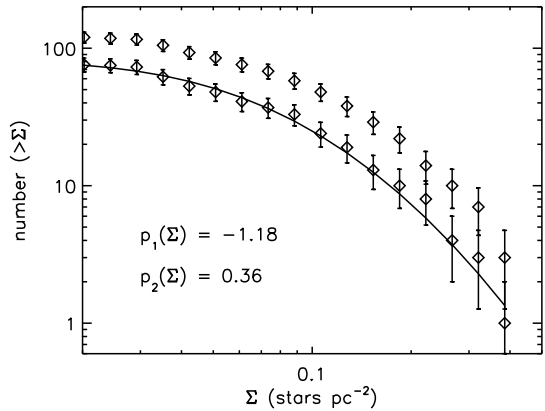}
\caption{Cumulative surface density distribution of young stellar structures. The upper data points are for all identified young stellar structures, which are affected by incompleteness; the lower ones are only for those with $R$~$>$~10~pc, the distribution of which is expected to well represent the distribution of the underlying young stellar structure population (see text). The solid line is the fit to the lower data points with a lognormal distribution. The error bars reflect Poissonian uncertainties.}
\label{density.fig}
\end{figure}

\begin{figure}
\centering
\includegraphics[scale=0.67,angle=0]{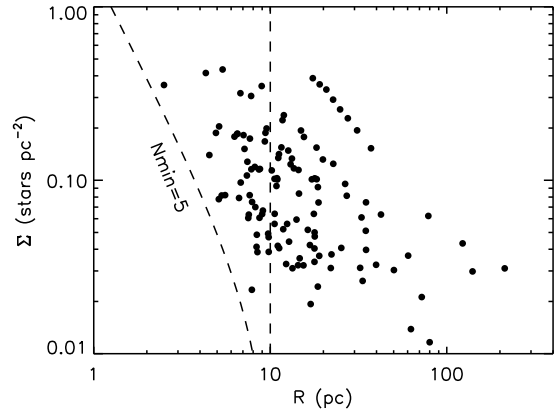}
\caption{Surface density--size relation of all identified young stellar structures. The vertical dashed line at $R$~=~10~pc indicates the resolution of the KDE map, and the tilted dashed line corresponds to the limit of $N_{\rm min}$~=~5 stars.}
\label{densitysize.fig}
\end{figure}

Figure~\ref{density.fig} shows the cumulative distribution of the structures' surface densities. The surface density (values given in Table~\ref{catalog.tab}) is calculated simply by the number of upper-MS stars divided by the contour area, i.e. $\Sigma$~=~$N_\ast$/A. As previously mentioned, the completeness of young stellar structures is affected by the resolution of the KDE map and the requirement of $N_{\rm min}$~=~5~stars for each structure. The relation between surface density and size is shown in Fig.~\ref{densitysize.fig}, in which the limits of the two constraits are also plotted. It is immediately obvious that the requirement of $N_{\rm min}$~=~5~stars rejects low-surface density structures at the small-size end. Beyond $R$~=~10~pc, however, this requirement does not affect the completeness (which reaffirms our conclusion in Section~\ref{size.sec}). On the other hand, the surface density does not show any apparent correlation with size. As a result, we expect that the surface density distribution for structures with $R$~$>$~10~pc should be a good representation of the distribution of the underlying young stellar structure population. Thus in summary, we can use structures with $R$~$>$~10~pc to investigate the distribution of the surface density. The result is shown as the lower data points in Fig.~\ref{density.fig}.

We further fit the data with
\begin{equation}
\label{density2.eq}
N(> \Sigma) = \dfrac{p_0}{2} \left[ 1 - {\rm erf}\left(\dfrac{{\rm ln} \Sigma - p_1}{p_2 \sqrt{2}} \right) \right],
\end{equation}
which is the cumulative form of the lognormal function
\begin{equation}
\label{density1.eq}
f(\Sigma) = \dfrac{p_0}{\Sigma p_2\sqrt{2\pi}} {\rm exp}\left[-\dfrac{({\rm ln} \Sigma - p_1)^2}{2p_2^2} \right],
\end{equation}
where $p_0$ is a normalization constant, $p_1$ the natural logarithm of the mean value, and $p_2$ the dispersion in e-foldings, respectively. The solid line in Fig.~\ref{density.fig} is the best-fitting result, which provides a reasonable description of the data. Consistent with this work, lognormal surface density distributions have also been reported by \citet{Bressert2010} for the young stellar objects in the solar neighborhood, and recently by \citet{Gouliermis2017} for the young stellar structures in the grand-design galaxy NGC~1566. Lognormal distributions of volume and/or column densities are also found for the ISM, either in hydrodynamical simulations of turbulent, isothermal gas \citep[e.g.][]{Klessen2000, Federrath2010, Konstandin2012} or in observations of molecular clouds \citep[e.g.][]{Lombardi2010, Schneider2012}.

The origin of lognormal density distributions can be understood in a purely statistical way in the context of hierarchical star formation with turbulence \citep{VazquezSemadeni1994}. In an ISM regulated by turbulence, a substructure with average density $\Sigma$ can be considered as hierarchically produced by an $n$-step sequence of density fluctuations, each occurring within a lower-density substructure whose average density is generated via fluctuation from the previous step. We denote this hierarchical sequence by \{$\Sigma_0$, $\Sigma_1$~=~$\epsilon_1\Sigma_0$, $\Sigma_2$~=~$\epsilon_2\Sigma_1$, ..., $\Sigma_{\rm n}$~=~$\epsilon_{\rm n}\Sigma_{\rm n-1}$\}, where $\epsilon_1$, $\epsilon_2$, ..., $\epsilon_{\rm n}$ would follow the same probability distribution function when pressure and self-gravity are unimportant, since in that case the hydrodynamic equations become self-similar and invariant to rescaling in density \citep[see the discussion of][their Section~2]{VazquezSemadeni1994}. As a result, $\Sigma$ is proportional to the product of this large number of identically-distributed random variables, and would follow a lognormal distribution according to the central limit theorem if the fluctuations are independent \citep[see also][their Section~3.3]{Federrath2010}. Furthermore, it is a natural expectation that the young stellar structures may have inherited the lognormal density distribution from their parental ISM substructures. Indeed, \citet{Gutermuth2011} report a correlation between the surface densities of young stellar objects and gas in eight nearby molecular clouds. The above theoretical considerations may be violated when the star-forming ISM is subject to strong self-gravity, shocks, rarefaction waves, etc., leading to non-Gaussian deviations from lognormal density distributions \citep{Klessen2000, Federrath2010, Girichidis2014}. The young stellar structures, however, do not demonstrate such deviations to any statistical significance.

\section{Temporal Evolution}
\label{evo.sec}

\subsection{The Upper-MS Subsamples}
\label{age.sec}

\begin{deluxetable*}{ccrccccc}
\tablecolumns{7}
\tablecaption{The upper-MS Subsamples. \label{sub.tab}}
\tablehead{
\colhead{Subsample} & \colhead{$K_{\rm s}$} & \colhead{$N_{\rm stars}$}
& \multicolumn{2}{c}{log($\tau$/yr)}
& \multicolumn{2}{c}{TPCF Slope} & $D_2 \tablenotemark{a}$ \\
\colhead{ } & \colhead{(mag)} & \colhead{ }
& \colhead{median} & \colhead{mean}
& \colhead{$<$~30~pc} & \colhead{30~--~300~pc} & }
\startdata
(a) & 14.5 -- 16.0 & 1298
& 7.2 & 7.2
& $-$0.50~$\pm$~0.06 & $-$0.25~$\pm$~0.01 & 1.75~$\pm$~0.01 \\
(b) & 16.0 -- 17.0 & 4733
& 7.4 & 7.5
& $-$0.29~$\pm$~0.08 & $-$0.17~$\pm$~0.02 & 1.83~$\pm$~0.02 \\
(c) & 17.0 -- 17.5 & 6108
& 7.6 & 7.7
& \multicolumn{2}{c}{$-$0.11~$\pm$~0.01} & 1.89~$\pm$~0.01 \\
(d) & 17.5 -- 18.0 & 13,123
& 8.0 & 7.9
& \multicolumn{2}{c}{$-$0.05~$\pm$~0.01} & 1.95~$\pm$~0.01
\enddata
\tablenotetext{a}{The fractal dimension is derived as $D_2$~=~$\eta$~+~2, where $\eta$ is the slope of the TPCF in the range of 30~$<$~$\theta$~$<$~300~pc.}
\end{deluxetable*}

\begin{figure*}
\centering
\begin{tabular}{cc}
\includegraphics[scale=0.67,angle=0]{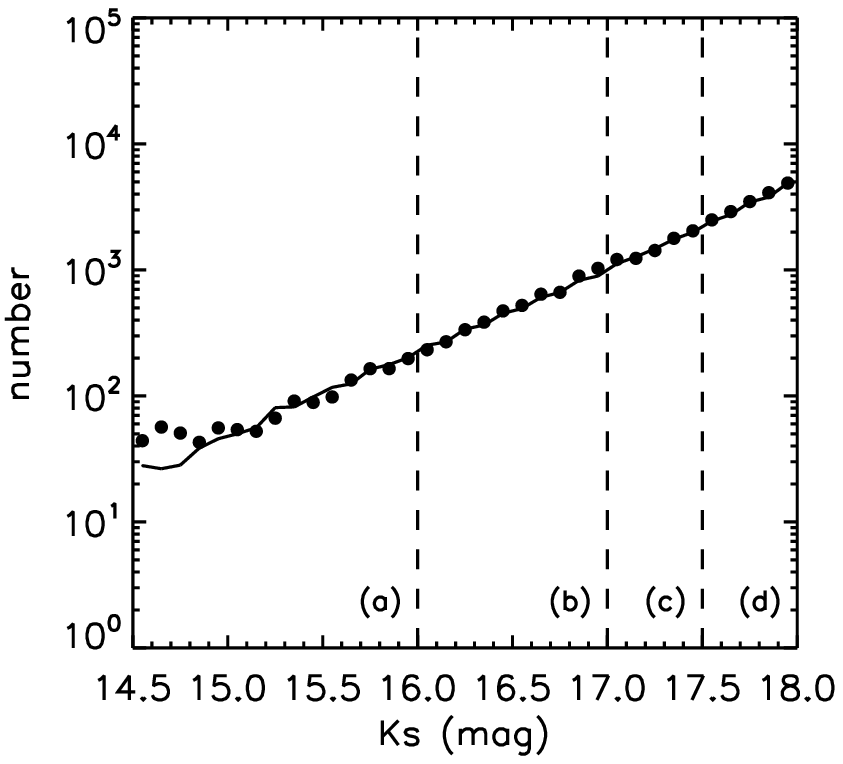}
\includegraphics[scale=0.67,angle=0]{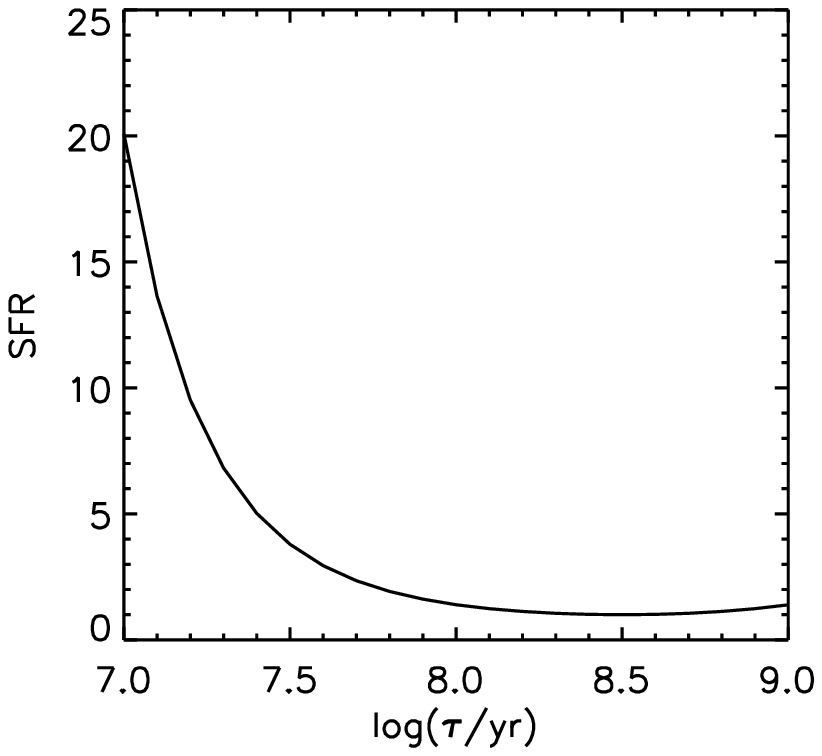}
\includegraphics[scale=0.67,angle=0]{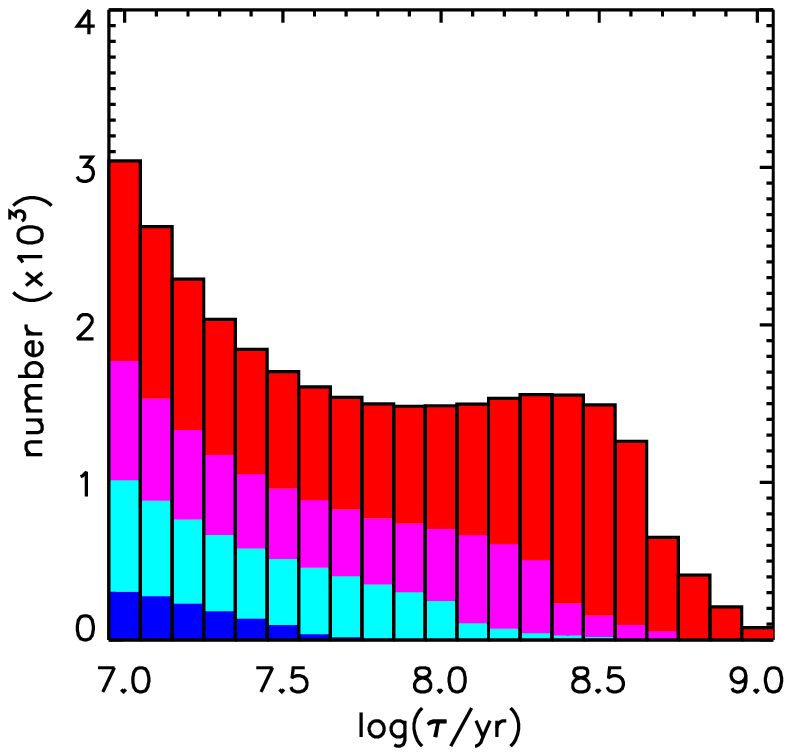}
\end{tabular}
\caption{Left: $K_{\rm s}$-band LFs of the observed and model upper-MS sample (points and solid line, respectively); the Poissonian error bars are comparable to or smaller than the symbol size. The vertical dashed lines indicate the magnitude cuts for the subsamples. Middle: the SFH adopted for modeling the upper-MS sample; the SFRs are given in arbitrary units. Right: age distribution of the simulated upper-MS stars. The colored areas correspond to stellar age distributions of the subsamples, with blue for subsample~(a), cyan for (b), magenta for (c), and red for (d).}
\label{sfr.fig}
\end{figure*}

The upper-MS sample contains stars of various ages, allowing us to investigate the temporal evolution of young stellar structures. To do this, we divide the full upper-MS sample into four subsamples (a), (b), (c), and (d), by applying three cuts in the $K_{\rm s}$-band magnitude at $K_{\rm s}$~=~16.0, 17.0, and 17.5~mag, respectively. The numbers of stars in the subsamples are given in Table~\ref{sub.tab}. While the brightest subsample~(a) consists of only the youngest stars, continuously fainter subsamples contain stars of wider age ranges. We estimate the stellar ages by constructing a model upper-MS sample which fits the luminosity function (LF) of the observed sample (Fig.~\ref{sfr.fig}, left-hand panel). In deriving the observed LF, we take into account photometric incompleteness by assigning weights to stars in the same way as in Section~\ref{mass.sec}.

The observed sample is assumed to be a linear combination of ``stellar partial models" (SPMs) -- model representations of stellar populations covering small age intervals \citep{Harris2001, Kerber2009, Rubele2012, Rubele2015}. Each SPM is simulated with PARSEC isochrones, the \citet{Chabrier2001} lognormal IMF, and a 30\% binary fraction. This binary fraction is typical for LMC star clusters \citep{Elson1998, Li2013}, and the simulated binaries are non-interacting systems with primary/secondary mass ratios evenly distributed from 0.7 to 1.0 (below these mass ratios, the secondary does not affect the photometry of the system significantly). The SPMs are reddened with an extinction value of $A_V$~=~0.6~mag as derived in Section~\ref{ums.sec}; we do not consider spatial variations of the extinction since the effect is small for $K_{\rm s}$-band magnitudes. Neither do we consider the photometric errors, typical values of which ($\sigma_{K_{\rm s}}$~=~0.02~mag at $K_{\rm s}$~=~18~mag) are much smaller than the LF's bin size (0.1~mag). The upper-MS stars of each SPM are selected based on the same criteria as have been applied to the data in Section~\ref{ums.sec}. We use 21 SPMs, each with an age interval of 0.1~dex in log($\tau$/yr) and central ages log($\tau$/yr)~=~7.0 to 9.0 in steps of 0.1~dex; and for simplicity, we assume that the 21 SPMs have a single metallicity of [Fe/H]~=~$-$0.3~dex, which is typical for massive MS stars in the LMC \citep[see e.g.][]{Rolleston2002}. Note that although the isochrone of a population of log($\tau$/yr)~=~9.0 does not cross the selection box for upper-MS stars (Fig.~\ref{cmd.fig}), its binary systems may still be bright enough to satisfy the selection criteria; thus, we choose log($\tau$/yr)~=~9.0 as the oldest age bin.

A linear combination of the SPMs corresponds to a certain star-forming history (SFH), which can be characterized by the star-forming rate (SFR) as a function of look-back time. Adopting different SFHs will produce model upper-MS samples with different LFs. Via tests with manually adjusted SFHs, we find that a SFH of the form
\begin{equation}
{\rm SFR} \propto {\rm exp}\left[\dfrac{({\rm log}(\tau/{\rm yr}) - 8.5)^2}{0.75}\right]
\end{equation}
(as shown in the middle panel of Fig.~\ref{sfr.fig}), provides a reasonable fit to the observed LF. The only large deviation exists in the three brightest magnitude bins; however, in the tests we have not considered the fitting performance here because there may be an increasing contribution from blue-loop stars (Section~\ref{ums.sec}). Adopting this SFH we are able to model the observed upper-MS sample and estimate the stellar age distributions in the full sample and subsamples, which are shown in the right-hand panel of Fig.~\ref{sfr.fig}. The full sample covers a wide age range from log($\tau$/yr)~=~7.0 to 9.0, with median and mean values of both 7.7. As expected, the subsamples, from the brightest to the faintest, cover progressively larger age ranges; their median and mean logarithmic stellar ages are given in Table~\ref{sub.tab}.

\subsection{Two-Point Correlation Functions}
\label{tpcf.sec}

\begin{figure*}
\centering
\begin{tabular}{cc}
\includegraphics[scale=0.60,angle=0]{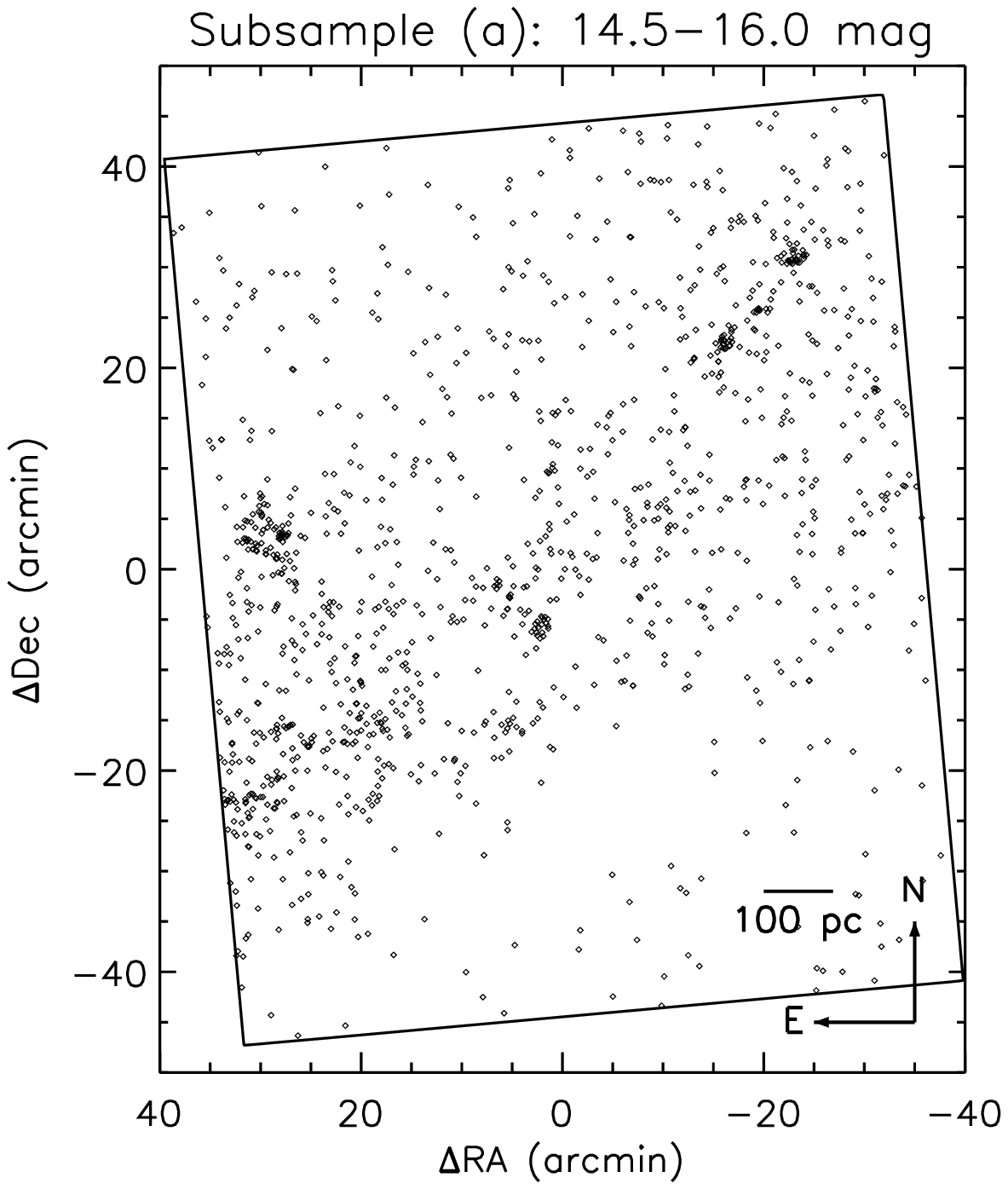}
\includegraphics[scale=0.60,angle=0]{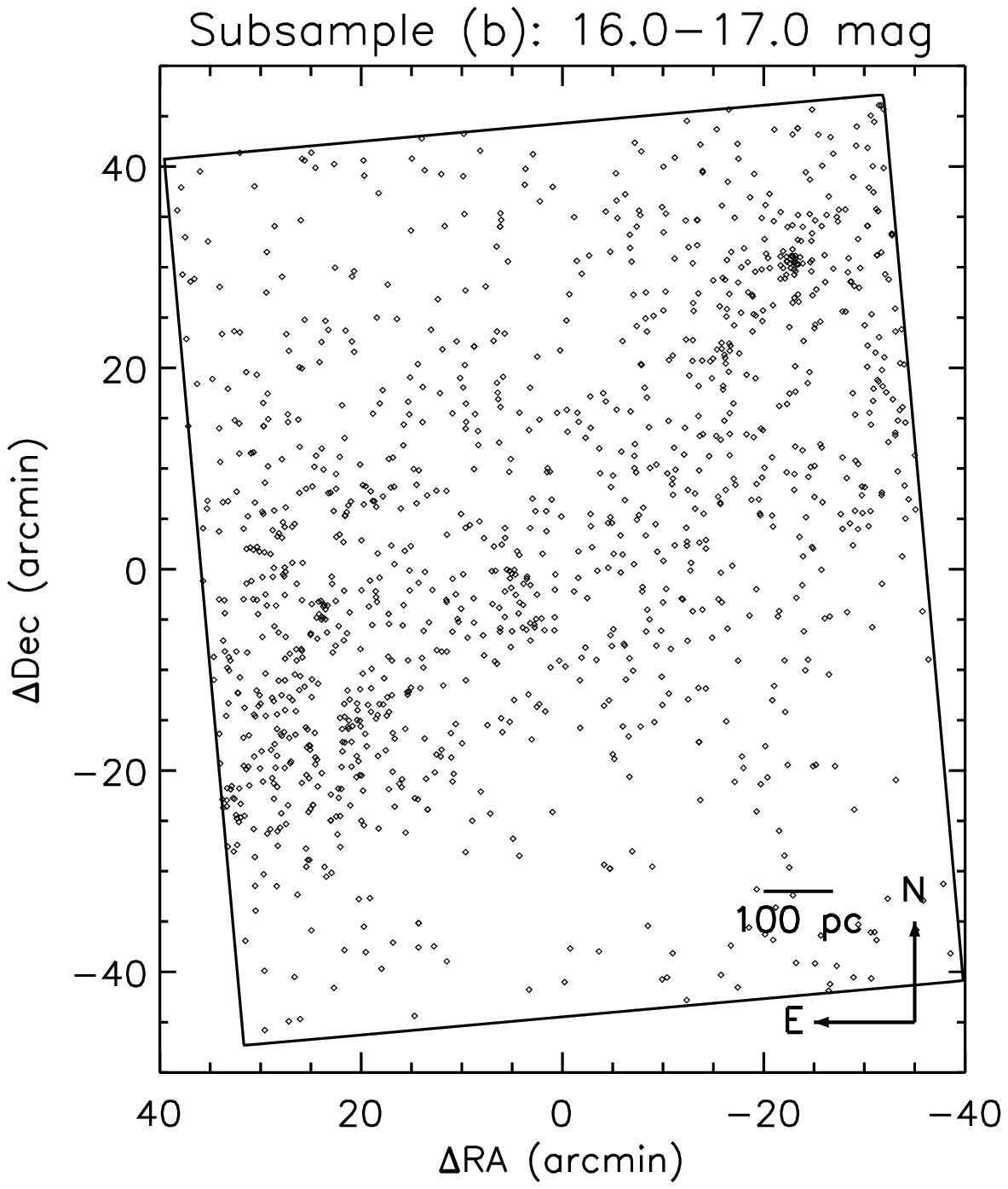} \\
\includegraphics[scale=0.60,angle=0]{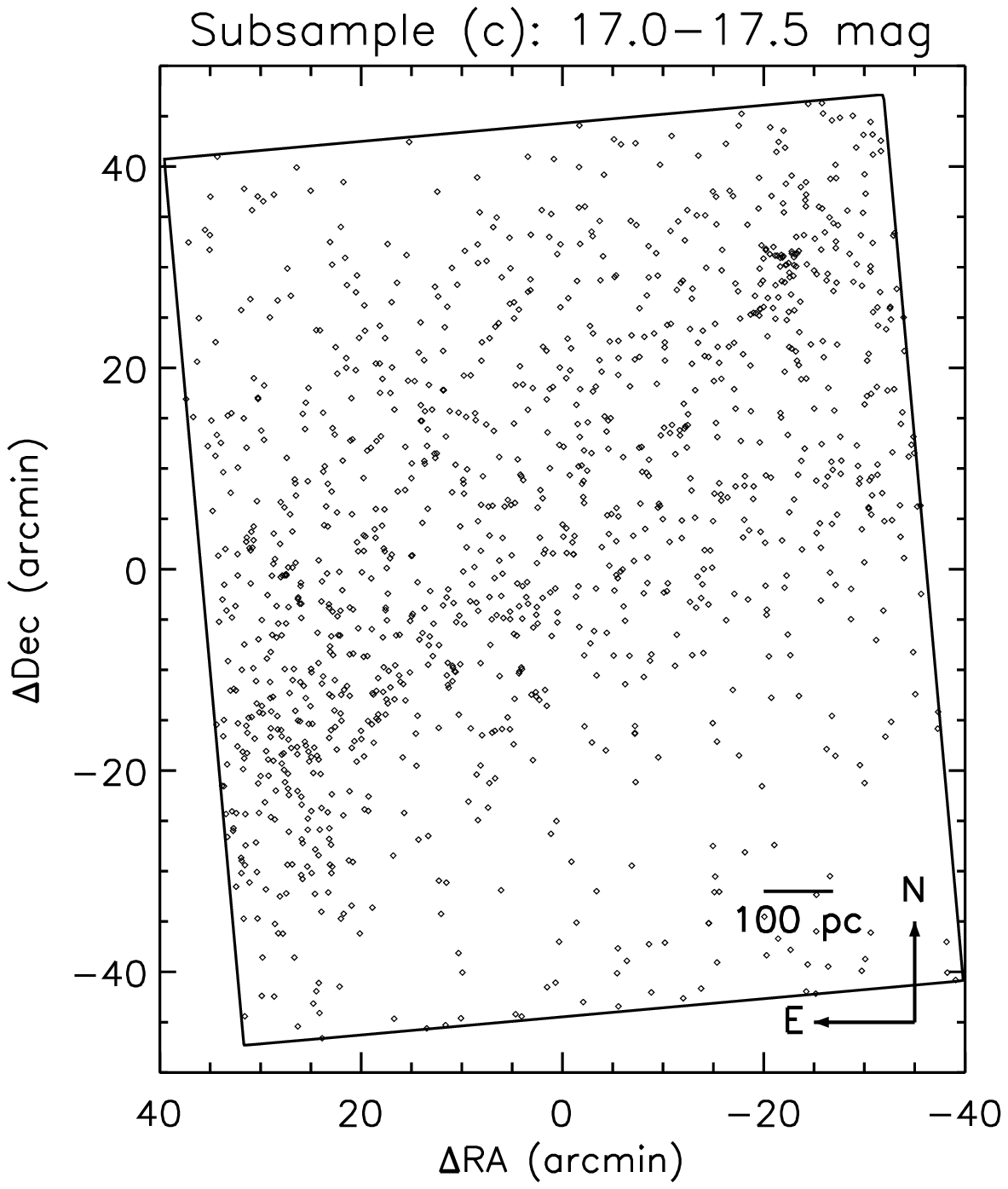}
\includegraphics[scale=0.60,angle=0]{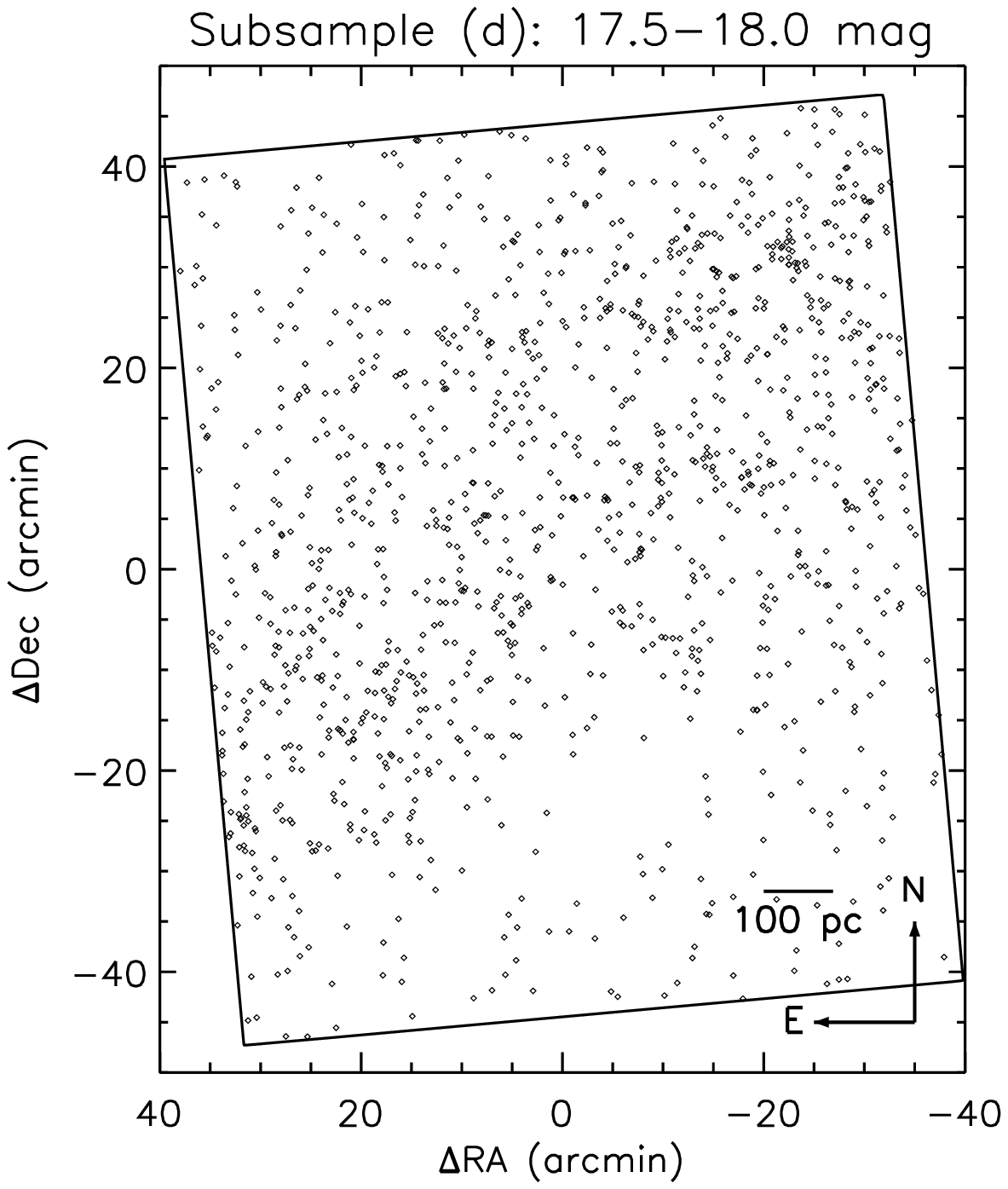} \\
\end{tabular}
\caption{Spatial distribution of upper-MS stars in the subsamples. In each panel, 1200 stars randomly selected from each subsample are displayed simply for clarity and for ease of comparison among subsamples. The (0, 0) position corresponds to R.A.(J2000)~=~05$^{\rm h}$12$^{\rm m}$55$^{\rm s}$.5, Dec.(J2000)~=~$-$69$^{\rm o}$16$\arcmin$41$\arcsec$, and the black rectangle shows the extent of VMC tile LMC~6\_4.}
\label{sub.fig}
\end{figure*}

Figure~\ref{sub.fig} shows the spatial distributions of upper-MS stars in the subsamples. It is apparent that subsample~(a) exhibits the most subclustered distribution. Subsample~(b) also displays a non-uniform distribution, but the stellar distribution is more dispersed and less subclustered. By contrast, subsamples~(c) and (d) have rather smooth stellar distributions. Visible in all four subsamples, the northeastern and southwestern areas always contain fewer stars than the central region; this is due to the large-scale density gradient of the bar structure (see also Fig.~\ref{lmc.fig} for the location of the bar with respect to the analyzed tile).

\begin{figure*}
\centering
\begin{tabular}{cc}
\includegraphics[scale=0.70,angle=0]{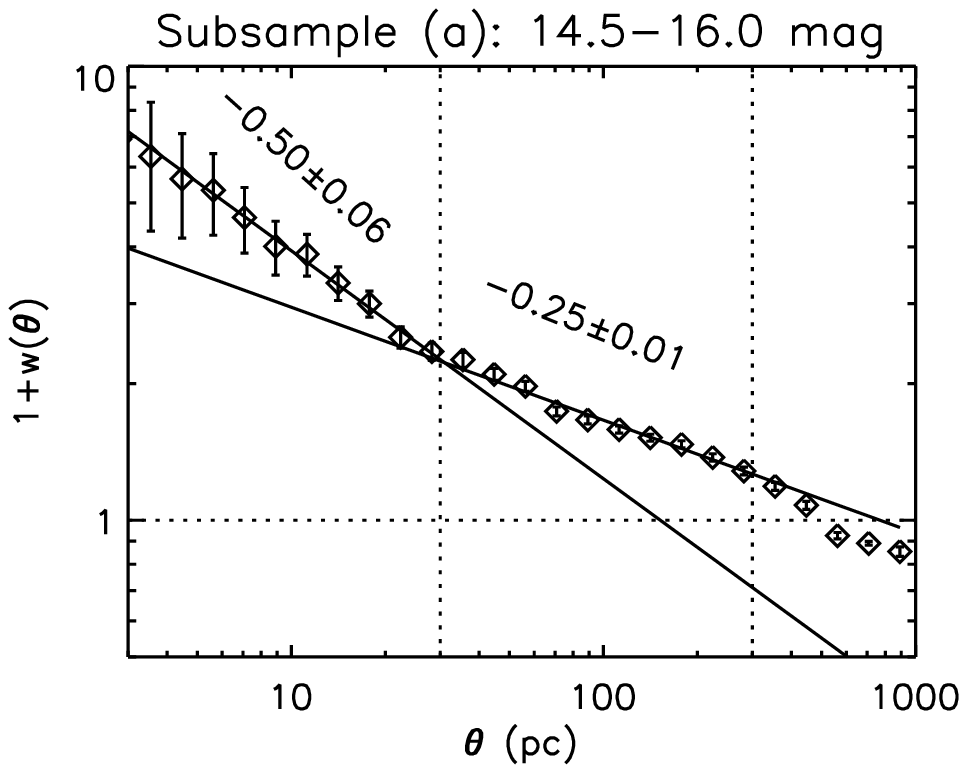}
\includegraphics[scale=0.70,angle=0]{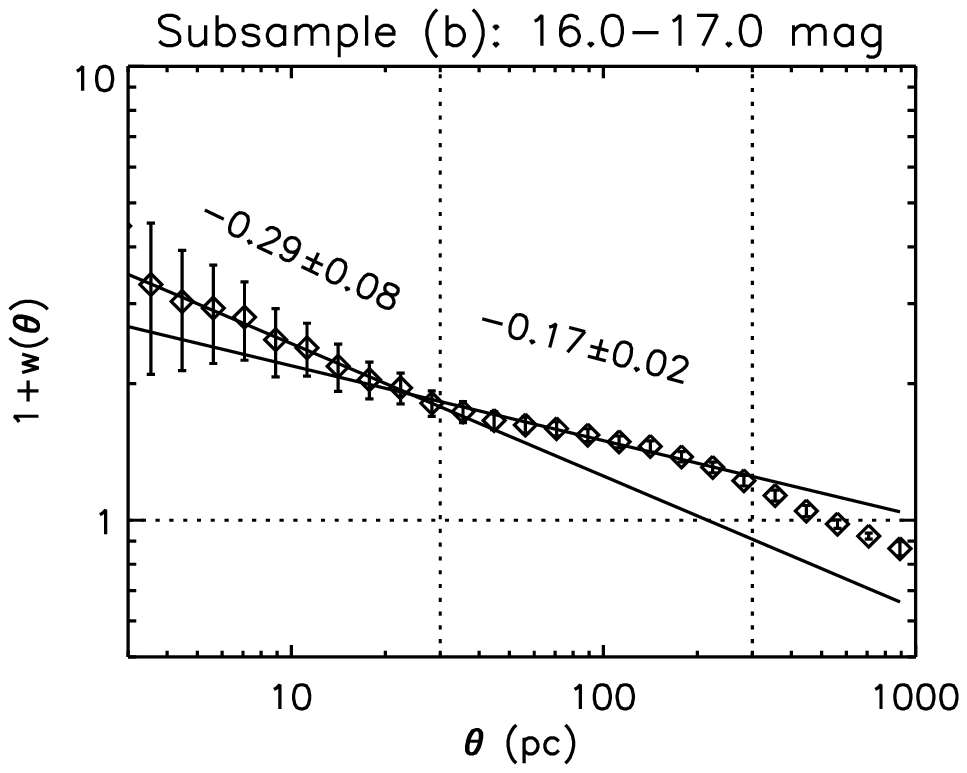} \\
\includegraphics[scale=0.70,angle=0]{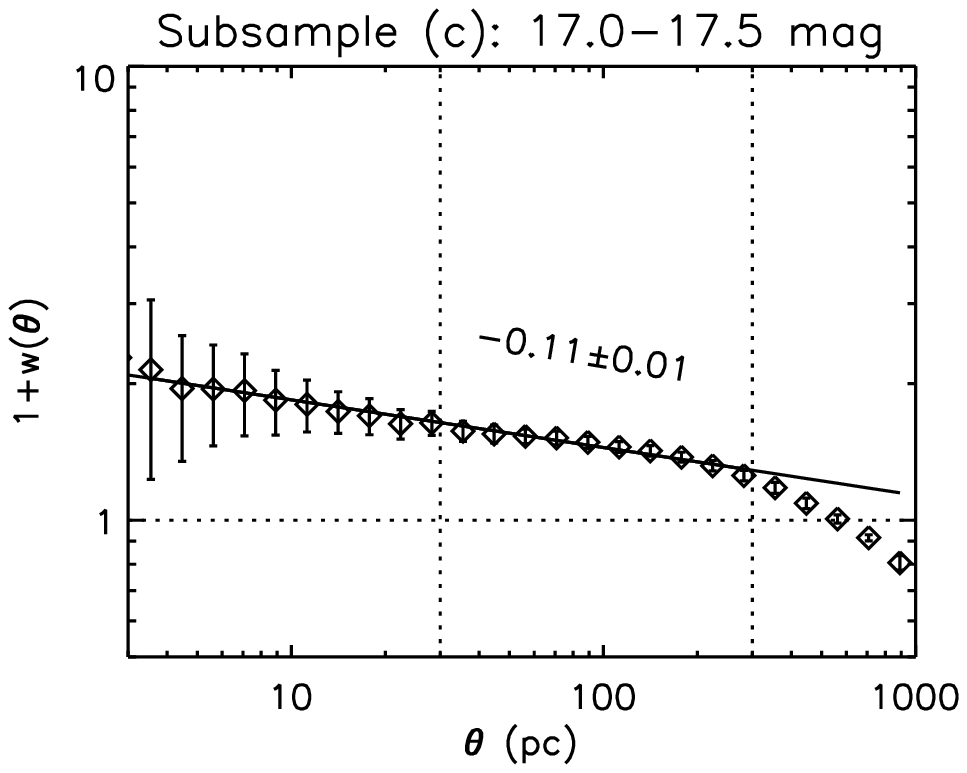}
\includegraphics[scale=0.70,angle=0]{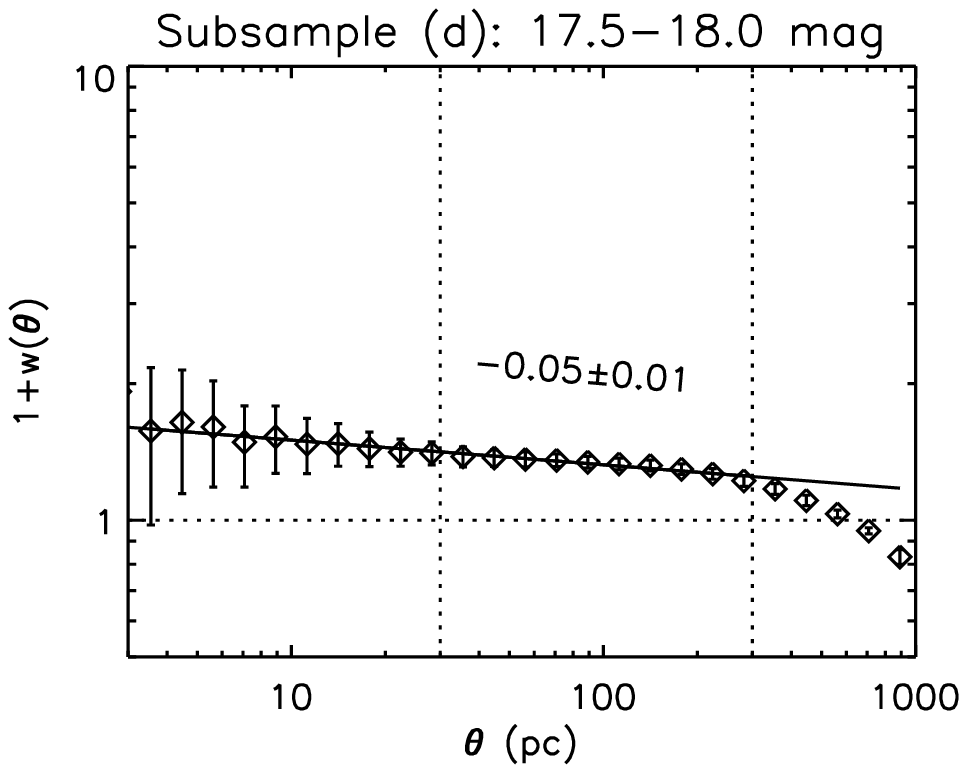} \\
\end{tabular}
\caption{TPCFs of the upper-MS subsamples. In each panel, the vertical dotted lines divide the large-, intermediate-, and small-separation regimes (see text), and the horizontal dotted line shows 1~+~$w(\theta)$~=~1 as expected for a random distribution without any substructures or density gradient. The solid line(s) are single or broken power-law fits to the data below $\theta$~$=$~300~pc, and the tilted numbers indicate their slopes.}
\label{tpcf.fig}
\end{figure*}

In the previous sections we have identified young stellar structures as surface overdensities from the KDE map (Section~\ref{kde.sec}). As will be discussed in Section~\ref{mix.sec}, however, structures identified in this way have a fractal dimension biased toward that of young stars. By contrast, the TPCF is a more suitable tool to quantify the distribution of stars of all ages in a sample. The TPCF, $w(\theta)$, is defined as the number of excess pairs of objects with a given separation, $\theta$, over the expected number for a reference distribution \citep{Peebles1980}. Here, we recast the TPCF as 1~+~$w(\theta)$, which is proportional to the more intuitive quantity of mean surface density of companions \citep{Larson1995, Kraus2008}. To save computing time and avoid edge effects, we evaluate the TPCFs via a Monte Carlo method. For each subsample, 1200 stars are randomly selected and the separation distribution of all their possible pairs, $N_p(\theta)$, is calculated; on the other hand, another 1200 artificial stars, which are randomly distributed across the tile, are generated as reference, and the separation distribution of all their possible pairs, $N_r(\theta)$, is derived. The TPCF from such a simulation is defined as 1~+~$w(\theta)$~=~$N_p(\theta)$/$N_r(\theta)$, and we repeat this process 1000 times; the final TPCF is obtained by averaging the results of all simulations, and their standard deviation is taken as the uncertainty in the TPCF.

The TPCFs of the four subsamples are shown in Fig.~\ref{tpcf.fig}. Beyond $\theta$~=~300~pc, all four subsamples show a steep drop in 1~+~$w(\theta)$, falling below unity at large separations.  Below $\theta$~=~300~pc, subsamples~(a) and (b) display broken power-law TPCFs, with steeper slopes at $\theta$~$<$~30~pc compared with shallower slopes for 30~$<$~$\theta$~$<$~300~pc; in contrast, the TPCFs of subsamples~(c) and (d) exhibit no statistically significant deviations from single power laws over the entire range of $\theta$~$<$~300~pc. The slopes are given in Table~\ref{sub.tab} and also indicated in the Fig~\ref{tpcf.fig}. The TPCFs in different separation regimes are discussed in detail in the following sections.

\subsubsection{The large-separation regime: effect of density gradient}
\label{large.sec}

In calculating the TPCFs, we have used a uniform distribution as reference. As previously mentioned, however, the bar causes a density gradient across the tile (Fig.~\ref{sub.fig}). The upper-MS stars are concentrated in a southeast--northwest locus, and there are fewer stars in the southwestern and northeastern regions. Thus, there are more large-separation stellar pairs in the reference than in the upper-MS subsamples. As a result, the TPCFs drop sharply beyond $\theta$~=~300~pc and fall below unity at even larger separations.

On the other hand, the density gradient occurs on a large scale comparable to the tile size, and is not significant on small scales. Thus, the small-scale clustering properties are still preserved in the TPCFs. For instance, the slopes of the TPCFs at $\theta$~$<$~300~pc, which may relate to the fractal dimensions and/or star clusters (see the next sections), should not be affected by the bar.

\subsubsection{The small-separation regime: effect of star clusters}
\label{small.sec}

\begin{figure*}
\centering
\begin{tabular}{cc}
\includegraphics[scale=0.60,angle=0]{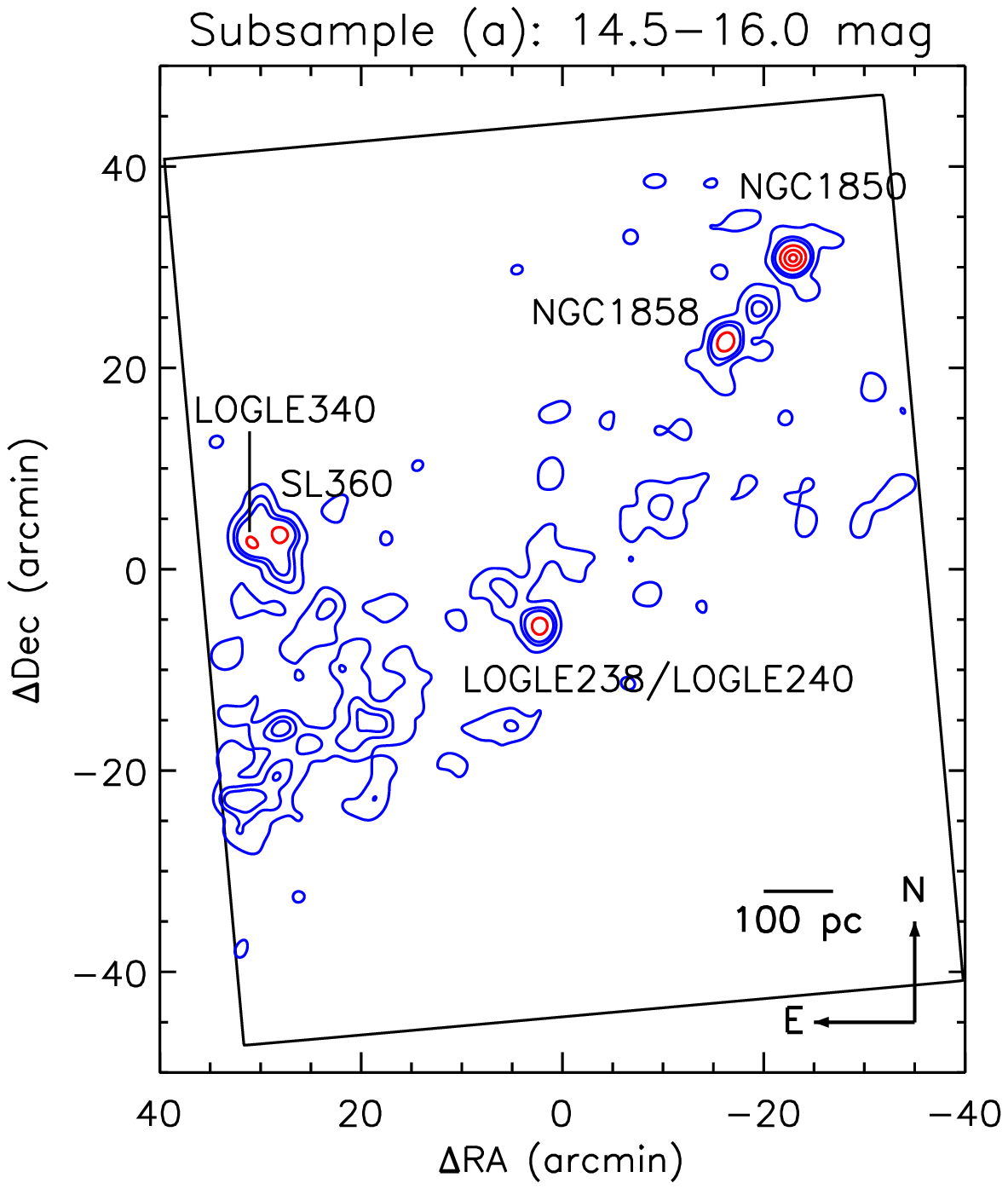}
\includegraphics[scale=0.60,angle=0]{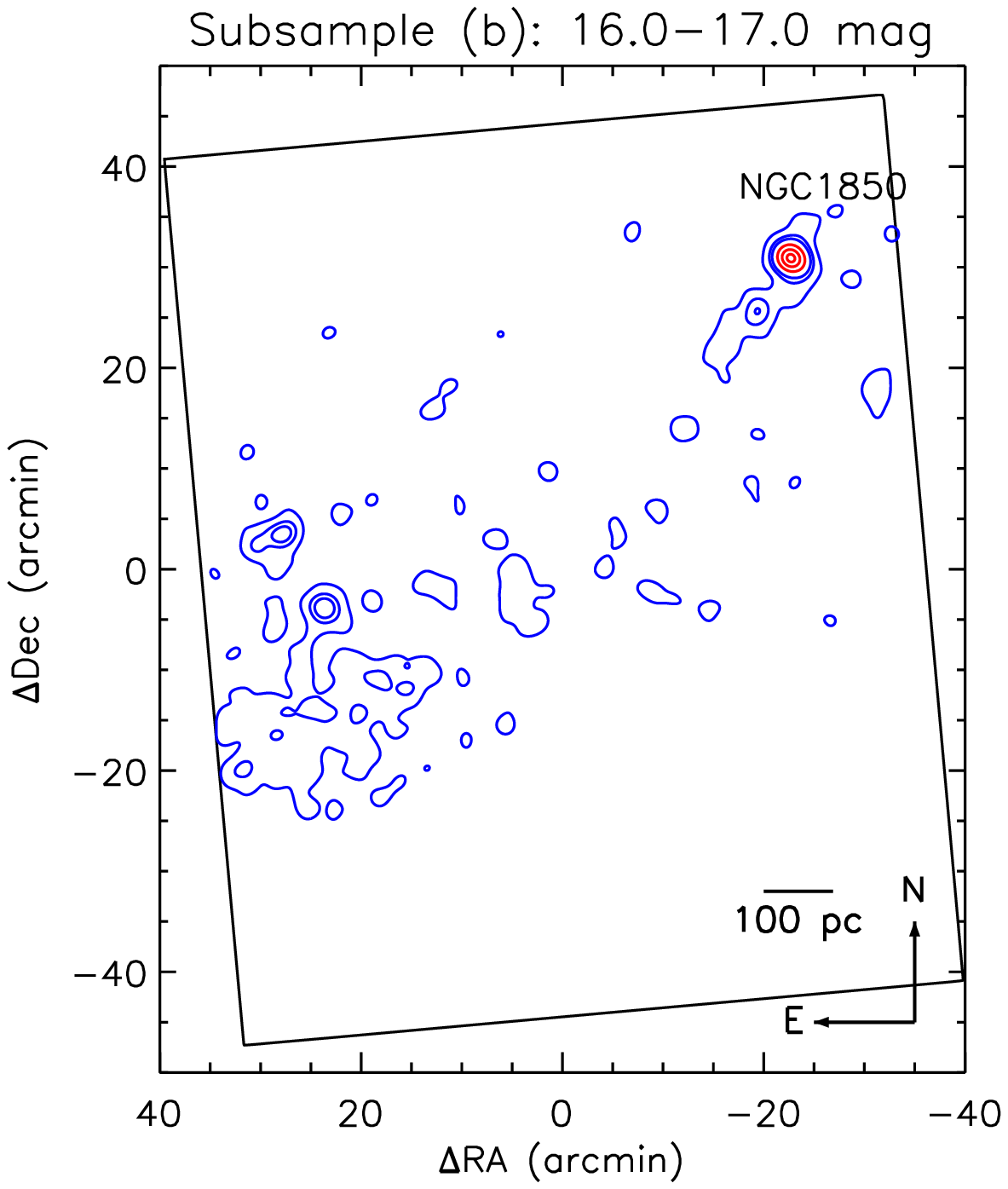}
\end{tabular}
\caption{Surface density maps of all possible stellar pairs with separation $\theta$~$<$~30~pc for subsamples~(a) and (b) (left- and right-hand panels, respectively). A stellar pair's position is represented by its center. The maps are convolved with a Gaussian kernel which is 10~pc in the standard deviation. The two maps have maximum surface densities of 0.78~pc$^{-2}$ and 3.49~pc$^{-2}$, respectively. The blue contours correspond to low-density levels of 1\%, 5\%, and 10\% of the maximum, while the red contours correspond to high-density levels of 30\%, 60\%, and 90\% of the maximum. The star clusters associated with the red, high-density contours are labeled in the figure. The (0, 0) position corresponds to R.A.(J2000)~=~05$^{\rm h}$12$^{\rm m}$55$^{\rm s}$.5, Dec.(J2000)~=~$-$69$^{\rm o}$16$\arcmin$41$\arcsec$, and the black rectangle shows the extent of VMC tile LMC~6\_4.}
\label{pair.fig}
\end{figure*}

We were curious about the break at $\theta$~=~30~pc in the TPCFs of subsamples~(a) and (b). To explore its origin, we show in Fig.~\ref{pair.fig} the spatial distributions of all possible pairs in the two subsamples with separations $\theta$~$<$~30~pc. In addition to some dispersed, low-density distributions, the stellar pairs are concentrated in one or several compact areas which are spatially associated with known star clusters. For both subsamples, the most significant concentration of stellar pairs is located in the northwest and associated with the populous star cluster NGC~1850. For subsample~(a), there are a few additional compact concentrations, whose associated star clusters have been labeled in the figure. Thus, it is reasonable to speculate that the break at $\theta$~=~30~pc is caused by the star clusters, whose member stars are so densely distributed, leading to an enhanced number of stellar pairs with $\theta$~$<$~30~pc than that expected from single power laws. The break is not seen in either subsample~(c) or (d), possibly because the numbers of non-cluster stars are so much larger than of the cluster stars; as a result, this effect is no longer significant.

Star clusters are subject to strong self-gravity and rapid dynamical evolutions. They are usually non-fractal, centrally concentrated objects, and reflect a different clustering mode from the fractal, hierarchical component \citep[see e.g.][]{Gouliermis2014}. Strictly speaking, star clusters can also be regarded as the high-density end of the continuous hierarchy of young stellar structures; moreover, young star clusters may in turn contain subclusters \citep{Gutermuth2005, Schmeja2008, Sanchez2009}. However, sophisticated simulations are needed to distinguish star clusters from the hierarchical component \citep[e.g.][]{Gouliermis2014}. On the other hand, stochastic sampling effects become increasingly important on small scales. Thus, we do not further explore the TPCFs at $\theta$~$<$~30~pc.

\subsubsection{The intermediate-separation regime: temporal evolution}
\label{intermediate.sec}

\begin{figure}
\centering
\includegraphics[scale=0.85,angle=0]{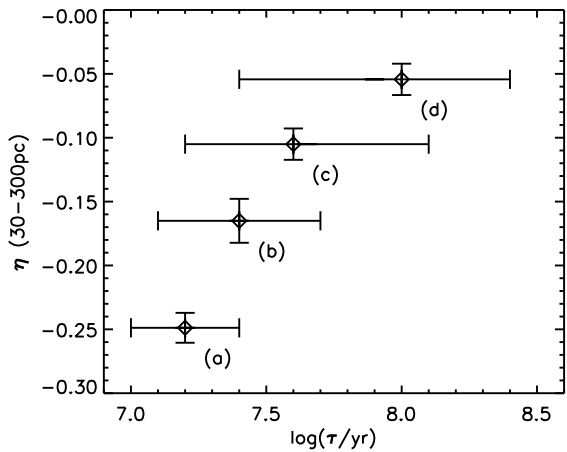}
\caption{TPCF slopes over 30--300~pc and median ages of the upper-MS subsamples. The horizontal erorr bars correspond to the interquartile ranges in stellar ages (i.e. between 25\% and 75\% observation).}
\label{wow.fig}
\end{figure}

In the range of 30~$<$~$\theta$~$<$~300~pc, the TPCFs of all four subsamples agree well with single power laws of the form 1~+~$w(\theta)$~$\propto$~$\theta^\eta$. The exponent is related to the projected fractal dimension as $D_2$~=~$\eta$~+~2 \citep{Larson1995, Kraus2008}, whose values are given in Table~\ref{sub.tab}. Smaller $\eta$ and $D_2$, corresponding to a steeper TPCF, suggest larger amounts of substructures, while larger values of $\eta$ and $D_2$, on the other hand, indicate a more uniform distribution. Fig.~\ref{wow.fig} shows the values of $\eta$ as a function of the median stellar age of the four subsamples. From subsample~(a) to (d), $\eta$ evolves steadily from $-$0.25 to $-$0.05, suggesting that they contain continuously fewer stellar structures as they become older. Subsample~(d) has $\eta$ very close to zero and is almost indistinguishable from a random distribution in this separation regime\footnote{Note that subsample~(d) is uniform only on small scales. On larger scales, it still exhibits a density gradient (Section~\ref{large.sec}). At $\theta$~$<$~300~pc, the TPCF is flat but above unity (Fig.~\ref{tpcf.fig}, bottom-right panel). This is because the upper-MS stars are more concentrated than the reference stars, which do not have any density gradient.}. Its median age, log($\tau$/yr)~=~8.0 or 100~Myr, provides a rough estimate of the timescale over which the young stellar structures disperse.

Temporal evolution of young stellar structures has also been investigated by \citet{Gieles2008}, \citet{Bastian2009}, and \citet{Gouliermis2015}. Specifically, \citet{Bastian2009} report a timescale of 175~Myr, or log($\tau$/yr)~=~8.2, for losing all substructures in the stellar distributions over the entire LMC. They note that this timescale is comparable to the dynamical crossing time of stars in this galaxy, suggesting that the dominant driver of the structures' evolution is general galactic dynamics. Dissolving star clusters may lead to the erasure of substructures; ejected stars from star clusters, e.g. runaway stars, may also account for the temporal evolution of the TPCFs \citep[see e.g.][]{Pellerin2007}. If these were the dominant mechanisms for the substructure dispersal, one would expect that the distribution of stars evolves more rapidly than that of star clusters, since in this case stars would have an additional expansion or ejection velocity component. However, \citet{Bastian2009} showed that the evolution of star clusters and stars is largely similar. Thus, although these effects may take place, they are not the dominant cause of the temporal evolution.

It still remains unanswered whether the structure dispersion timescale may vary with location in the galaxy, as the young stellar structures may reside in different environments and undergo different physical processes. For instance, the bar complex is expected to be influenced by bar perturbations, if any (\citealt{Gardiner1998}; see also the discussion against a dynamical bar in Section~5.1.1 of \citealt{Harris2009}). Compared with the galaxy outskirts, it is less affected by the tidal forces from the SMC or the Milky Way \citep[e.g.][]{Fujimoto1990, Bekki2007, DOnghia2016}. It is thus interesting to compare the structure dispersion timescales for the global LMC \citep[175~Myr from][]{Bastian2009} and for the bar complex only (100~Myr).

However, we note that the dispersion timescale is subject to significant uncertainties. On the one hand, the derived median ages may vary with the adopted SFH and metallicity of the SPMs (Section~\ref{age.sec}). We redid the calculation in Section~\ref{age.sec} by changing the metallicity to [Fe/H]~=~$-$0.5 or 0.0~dex (the SFH changes accordingly to fit the observed LF). The derived median ages of the subsamples change by at most 0.1~dex, suggesting that the effect of this change is unimportant. On the other hand, the largest uncertainty comes from the significant spread of stellar ages within a subsample. In particular, while the brighter subsamples consist of only young stars, the fainter ones have larger age spreads because they contain both young and old stars. The interquartile range of log($\tau$/yr) offers an estimate of the age spread. For instance, the interquartile range of subsample~(d) is as large as [7.4~dex, 8.4~dex]. This is a very large spread, even exceeding the difference between the median ages of the subsamples. In other words, the subsamples are not single-aged stellar populations, and the TPCFs reflect the mixed distribution of stars of different ages. Thus, it is not easy to obtain an accurate measurement of the structure dispersion timescale with upper-MS stars. Considering this fact, the timescale of 100~Myr as derived here is not inconsistent with the value from \citet{Bastian2009}, which is 175~Myr for the whole LMC.

\subsection{Fractal Dimension from Mixed Populations}
\label{mix.sec}

In this section we have demonstrated that young stellar structures evolve rapidly after birth and are finally dispersed on a timescale of $\sim$100~Myr. We have also shown that the upper-MS stars cover a wide age range, with a significant fraction older than 100~Myr. Still, a small fractal dimension of $D_2$~=~1.5~$\pm$~0.1 was previously derived from the structures' size distribution and mass--size relation. This is very close to that expected for the ISM and newly formed stars ($\sim$1.4; see Section~\ref{size.sec}), and seems contradictory to the presence of old stars in the upper-MS sample. We recall, however, that the young stellar structures in Section~\ref{yss.sec} were identified as surface overdensities from the KDE map. Given that younger populations are more substructured, the overdensities are dominated by the young stars in the upper-MS sample. Slightly older stars make minor contributions, and very old stars lead to density enhancements only through their statistical fluctuations. Thus, derivation of small fractal dimension is expected.

\begin{figure}
\centering
\includegraphics[scale=0.70,angle=0]{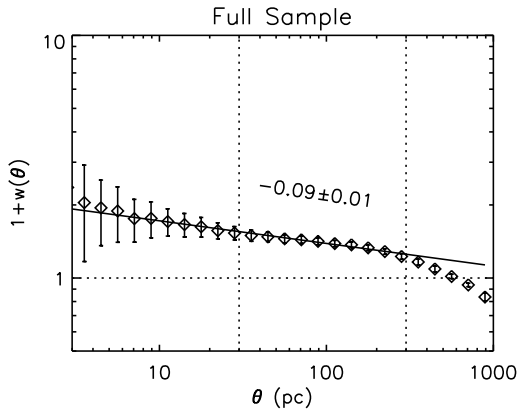}
\caption{As Fig.~\ref{tpcf.fig} but for the full upper-MS sample.}
\label{xcf.fig}
\end{figure}

The fractal dimension can also be derived from the TPCF (e.g. Section~\ref{tpcf.sec}), to which, however, both young and old stars contribute importantly. With an increasing number of old stars, the relative number of small-separation stellar pairs decreases, leading to shallower TPCFs and larger fractal dimensions. Thus, even with the same sample of stars, the fractal dimension from the TPCF is often higher than that from surface overdensities, as long as the sample has a significant age spread.

To verify this, we calculate the TPCF for the full upper-MS sample (Fig.~\ref{xcf.fig}) in the same way as for the subsamples. The full-sample TPCF is also well described by a single power law below $\theta$~=~300~pc, the slope of which is $\eta$~=~$-$0.09~$\pm$~0.01. This suggests a fractal dimension of $D_2$~=~1.91~$\pm$~0.01, significantly larger than that from the size distribution or mass--size relation. Actually, this value is very close to the TPCF-derived fractal dimensions of subsamples~(c) and (d), since these two subsamples make up the majority of stars in the full sample (Table~\ref{sub.tab}).

From this perspective, we can also understand the fractal dimension of subsample~(a) derived with the TPCF (1.75~$\pm$~0.01, Table~\ref{sub.tab}). Although this subsample has a young median age, it also contains many old stars up to $\sim$30~Myr (Figs.~\ref{sfr.fig} and \ref{wow.fig}). Within this timescale, the young stellar structures have already undergone significant evolution. Thus, its fractal dimension from the TPCF is significatly higher than $\sim$1.4, which is expected for the ISM and newly formed stars.

\section{Summary and Conclusions}
\label{summary.sec}

In this paper, the star-forming complex at the northwestern end of the LMC bar region (the bar complex) has been investigated with the VMC survey. The analysis is based on $\sim$2.5~$\times$~10$^4$ upper-MS stars selected from the CMD, which is an order of magnitude larger than the samples used in previous studies. As a result, we have been able to trace young stellar structures down to scales of $\sim$10~pc.

The upper-MS stars have highly non-uniform spatial distributions. Young stellar structures in the complex are identified as stellar surface overdensities from the KDE map at different significance levels. The structures are organized in a hierarchical way such that larger, lower-level structures host one or several smaller, higher-level ones inside. This ``parent--child" relation is further illustrated with dendrograms for two typical young stellar structures along with their substructures.

Size, mass, and surface density are also calculated for the young stellar structures. In the range not affected by incompleteness, the size distribution can be well described by a single power law, which is consistent with a projected fractal dimension $D_2$~=~1.5~$\pm$~0.1. The structures follow a power-law mass function of the form $n(M)$d$M$~$\propto$~$M^{-\beta}$d$M$, with $\beta$~=~2.0~$\pm$~0.1. Their surface densities, on the other hand, agree well with a lognormal distribution. These results support the scenario of hierarchical star formation regulated by turbulence, in which newly-formed stars follow the gas distribution. The effects of other physical processes, e.g. gravity, strong shocks, rarefaction waves, and environmental influences, are not obviously visible from our results.

We further divide the full upper-MS sample into subsamples (a--d) with different magnitude ranges; the subsamples have increasing average ages, as confirmed by detailed LF modeling. The stellar distributions of the subsamples are quantitively studied based on the TPCFs, where the presence of the bar and star clusters is also apparent beyond 300~pc and below 30~pc, respectively. In the spatial range of 30--300~pc, the TPCFs are well described by single power laws, the slopes of which increase continuously from subsamples~(a) to (d). Over this separation range, the youngest subsample~(a) contains most substructures, while the oldest subsample~(d) is almost indistinguishable from a uniform distribution. This suggests rapid temporal evolution so that the young stellar structures are completely dispersed on a timescale of $\sim$100~Myr. Considering the uncertainties, however, this timescale is not significantly different from that previously reported for the entire LMC. Thus, the environmental influence from the bar, if any, is not revealed from the data to any statistical significance.

Considering this evolutionary effect, we further point out that the fractal dimension may be method-dependent even using the same sample, as long as the stellar sample have a significant age spread. $D_2$ obtained with surface overdensities is biased toward that of young stars. On the other hand, $D_2$ from the TPCF is also sensitive to the old stars. As a result, $D_2$ will often appear lower if it is derived with the former method.

\acknowledgements

We thank the anonymous referee for the helpful suggestion on this paper. The analysis in this paper is based on observations collected at the European Organisation for Astronomical Research in the Southern Hemisphere under ESO program 179-B-2003. We thank the Cambridge Astronomy Survey Unit (CASU) and the Wide Field Astronomy Unit (WFAU) in Edinburgh for providing calibrated data products through the support of the Science and Technology Facility Council (STFC) in the UK. We thank Jim P. Emerson for his suggestions. \text{N.-C.~S} and \text{R.~d.~G.} acknowledge funding support from the National Natural Science Foundation of China through grants 11373010, 11633005, and U1631102. \text{S.~S.} acknowledges research funding support from Chinese Postdoctoral Science Foundation (grant number 2016M590013). \text{M.-R.~L.~C.} acknowledges support from the STFC (grant number ST/M001008/1).


\end{document}